\documentclass[reprint,amsmath,amssymb,
 aps]{revtex4-2}

\usepackage{graphicx}
\usepackage{dcolumn}
\usepackage{bm,ulem}
\usepackage[pdftex,
            colorlinks=true,
            citecolor=blue,
            linkcolor=blue]{hyperref}
\usepackage{appendix}

\begin{document}

\preprint{APS/123-QED}

\title{Quasiclassical theory of vortex states in locally non-centrosymmetric superconductors: application to $\mathrm{CeRh_2As_2}$}

\author{Akihiro Minamide}
\email{minamide.akihiro.34n@st.kyoto-u.ac.jp}
\affiliation{Department of Physics, Kyoto University, Kyoto 606-8502, Japan}

\author{Youichi Yanase}
\affiliation{Department of Physics, Kyoto University, Kyoto 606-8502, Japan}

\date{\today}

\begin{abstract}
$\mathrm{CeRh_2As_2}$, a heavy fermion superconductor discovered in 2021, exhibits two distinct superconducting phases under a $c$-axis magnetic field.
This unconventional phase diagram has been attributed to the local inversion symmetry breaking at the $\mathrm{Ce}$ sites.
At low magnetic fields, a conventional even-parity spin-singlet superconducting state is realized, whereas at higher fields, an odd-parity spin-singlet superconducting state, in which the order parameter alternates sign between neighboring Ce layers, becomes stabilized.
In this study, we employ a quasiclassical approach to investigate the vortex states of bilayer superconductors with locally broken inversion symmetry. 
We calculate the phase diagram and the local density of states (LDOS) in the vortex lattice state and find that the pairing symmetry of different superconducting states is clearly manifested in the peak structure of LDOS at the vortex core.
Since LDOS is experimentally observable, our work provides a pathway for experimental verification of the superconducting parity transition in $\mathrm{CeRh_2As_2}$.
\end{abstract}

\maketitle

\section{Introduction}
\label{sec:introduction}
A recently-discovered heavy fermion superconductor $\mathrm{CeRh_2As_2}$~\cite{khim2021field} has received a great deal of attention as a new member of the multiphase superconductor family.
Under a magnetic field directed along the $c$ axis, it exhibits a thermodynamic phase transition around $\mu_0 H^*=4\,\mathrm{T}$, which separates different superconducting phases~\cite{khim2021field,landaeta2022field,semeniuk2023decoupling,khanenko2025phase}.
There are several possible scenarios to explain this multiphase superconductivity~\cite{mockli2021two,machida2022violation,hazra2023triplet,szabo2024superconductivity}, the most persuasive of which explains it in terms of local inversion symmetry breaking of the crystal structure~\cite{yoshida2012pair,fischer2023superconductivity}.
The crystal structure of $\mathrm{CeRh_2As_2}$ has a global inversion center, but the $\mathrm{Ce}$ sublayers locally lack inversion symmetry because they are sandwiched between two inequivalent blocks composed of $\mathrm{Rh}$ and $\mathrm{As}$ atoms.
In such a locally noncentrosymmetric system, the superconducting order parameter acquires an additional degree of freedom, the sublattice.
When the magnetic field is applied perpendicular to the layers, the BCS state with even-parity spin-singlet order parameter is stable at low fields, while the pair-density-wave (PDW) state with odd-parity spin-singlet order parameter is stabilized at high fields~\cite{maruyama2012locally,yoshida2012pair}.
This theoretically predicted phase transition is called the parity transition, and $\mathrm{CeRh_2As_2}$ is considered to be the first material in which the superconducting parity transition was observed.

The appearance of the PDW state has also been predicted for locally noncentrosymmetric crystals other than the layered system~\cite{fischer2011superconductivity,watanabe2015odd,sumita2016superconductivity,nakamura2017odd}.
However, the fact that the parity transition has been observed only in $\mathrm{CeRh_2As_2}$ implies a peculiarity of this material, which enhances the importance of further studies on it.
Here, we review some features of CeRh$_2$As$_2$.
(1) The distinctive features of $\mathrm{CeRh_2As_2}$ include the strong correlation effect originating from $\mathrm{Ce}\ 4f$ electrons.
The resistivity shows the Kondo effect with $T_{\rm K}\sim 30\,\mathrm{K}$, and the specific heat exhibits non-Fermi liquid behaviors, indicating the proximity to a quantum critical point~\cite{khim2021field,hafner2022possible,khanenko2025origin,pfeiffer2024pressure}.
Consistent with experimental indication, a theory that incorporates strong correlation effects yields a better reproduction of the experimental phase diagram than weak-coupling calculations~\cite{nogaki2022even}.
(2) For the band structure, angle-resolved photoemission spectroscopy (ARPES) measurements~\cite{wu2024fermi,chen2024exploring,chen2024coexistence} and first-principles band calculations~\cite{ishizuka2024correlation,nogaki2021topological,ptok2021electronic} have revealed the rare coexistence of the van Hove singularity and the $4f$-electron flat bands, which implies unique correlation effects in this material.
(3) The presence of additional correlated electron phases is also notable.
At zero magnetic field, an anomaly of specific heat was observed at $T_0=0.4\,\mathrm{K}$ above the superconducting transition temperature $T_{\mathrm{SC}}=0.26\,\mathrm{K}$~\cite{khim2021field}.
This anomaly was initially interpreted as the onset of a non-magnetic quadrupole-density-wave order, since the bulk magnetic probe did not capture the corresponding signal in early stage samples~\cite{hafner2022possible,mishra2022anisotropic,landaeta2022field,semeniuk2023decoupling}.
However, recent specific heat~\cite{chajewski2024discovery} and muon spin relaxation ($\mu$SR) experiments~\cite{khim2025coexistence} have revealed the magnetic characteristics of this ordered phase, suggesting a connection to the antiferromagnetic order reported in the nuclear quadrupole resonance (NQR)~\cite{kibune2022observation} and nuclear magnetic resonance (NMR) studies~\cite{kitagawa2022two,ogata2023parity,ogata2023investigation,ogata2024appearance}.
Consistently, antiferromagnetic fluctuation has been observed in the inelastic neutron scattering experiment~\cite{chen2024quasi} and has been attributed to the nesting properties of the Fermi surface~\cite{chen2024coexistence,wu2024fermi,chen2024exploring}.
In addition, magnetostriction and ac-susceptibility experiments~\cite{khanenko2025phase} have clarified that this ordered phase coexists with the low-field superconducting phase and a part of the high-field superconducting phase.
Various theoretical studies have been devoted to unraveling this enigmatic order~\cite{thalmeier2025thermodynamics,miyake2024possible,schmidt2024anisotropic,jakubczyk2025composite,lee2025unified}, and several works offer alternative scenarios for the multiple superconducting phases in terms of the interplay of superconductivity and other correlated orders~\cite{machida2022violation,szabo2024superconductivity}.
However, the experimentally observed phase diagram under pressure suggests negligible coupling between superconductivity and other correlated phases~\cite{pfeiffer2023exposing}.

Theoretical studies on locally noncentrosymmetric superconductors have mainly focused on spatially homogeneous states~\cite{yoshida2012pair,yoshida2014parity,sigrist2014superconductors,nogaki2022even,nogaki2024field,fischer2011superconductivity,watanabe2015odd,sumita2016superconductivity,nakamura2017odd,lee2025unified,mockli2021superconductivity,nally2024phase,amin2024kramers,maruyama2012locally,fischer2023superconductivity} under the assumption of the Pauli-limited regime, i.e., the Maki parameter $\alpha_{\mathrm{M}}$~\cite{maki1966effect} is infinity.
This assumption has been justified on the grounds that the large effective mass of quasiparticles in $\mathrm{CeRh_2As_2}$ suppresses the orbital depairing effect.
Such approaches have successfully grasped the physics of $\mathrm{CeRh_2As_2}$.
Nevertheless, in principle, consideration of the vortex degrees of freedom is required to study type-II superconductors subjected to a magnetic field.
In fact, the Maki parameter typically takes a moderate value $\alpha_{\mathrm{M}}=3\sim 5$ in heavy fermion superconductors~\cite{matsuda2007fulde}.
Particularly, in $\mathrm{CeRh_2As_2}$, understanding of the vortex lattice states is essential because superconductivity survives up to considerably high magnetic fields~\cite{khim2021field}, where the inter-vortex spacing becomes short and the overlap between neighboring vortices is non-negligible.
Only a few previous studies have addressed the effect of vortices explicitly.
Higashi \textit{et al.}~\cite{higashi2016robust} dealt with a single isolated vortex based on the quasiclassical Eilenberger formalism, and M\"{o}ckli \textit{et al.}~\cite{mockli2018orbitally} used the phenomenological Ginzburg-Landau (GL) model with the circular cell method to analyze the mixed state.
Although both studies elucidated the characteristics of the vortex, each failed to take into account some aspects of the vortex states: the former did not discuss inter-vortex interactions and the thermodynamical stability of vortex states, while the latter lacks the microscopic foundation for its GL free energy functional.
In our previous work~\cite{minamide2025superconducting}, the GL model of vortex states was derived microscopically from the bilayer Rashba model and, as a result, the $H$-$T$ phase diagram of $\mathrm{CeRh_2As_2}$ was reproduced quantitatively.
Moreover, we predicted that the third superconducting state, which we named the superconducting meron state, stabilizes near the multicritical point.
However, since the GL theory is valid only around the transition temperature, it remains insufficient for the analysis of vortex states over the entire region of the superconducting phase.

In this paper, we investigate the vortex lattice state of locally noncentrosymmetric superconductors by applying the quasiclassical theory~\cite{eilenberger1968transformation,larkin1969quasiclassical} to the bilayer Rashba model.
The thermodynamic stability of the BCS and PDW states is evaluated and consequently the characteristic $H$-$T$ phase diagram is determined down to the low temperature region.
Also, the local density of states is calculated, which manifests the electronic bound states around a vortex core in each superconducting state.

The remainder of this paper is structured as follows.
In Sec.~\ref{sec:model_and_method}, we present the model Hamiltonian and derive the multiband quasiclassical theory.
In Sec.~\ref{sec:results_and_discussion}, we calculate the free energy of superconducting states down to the low-temperature regime and also investigate the spatially-resolved nature of the vortex cores.
Finally, Sec.~\ref{sec:conclusion} summarizes the results and discusses related experiments that would validate the scenario of a superconducting parity transition in $\mathrm{CeRh_2As_2}$.
Throughout this paper, we use the unit $\hbar=c=k_{\mathrm{B}}=1$, and $e(>0)$ denotes the elementary charge.

\section{Model and Method}
\label{sec:model_and_method}
\begin{figure}[tbp]
  \centering
  \includegraphics[width=0.30\textwidth]{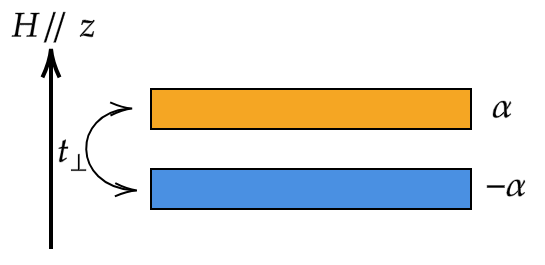}
  \caption{Schematic figure of the bilayer Rashba model.
  This model supposes the bilayer system, in which the inversion symmetry is locally broken in each layer.
  Interlayer hopping $t_\perp$, external magnetic field $H$, and layer-dependent ASOC $\alpha$ upon these layers are illustrated.
  }
\label{fig:bilayer_system_schematic}
\end{figure}

In this paper, $\check{a}$, $\hat{a}$, and $\tilde{a}$ represent a $8\times 8$, $4\times 4$, and $2\times 2$ matrix, respectively.
Let $c_{sl}(\bm{r})\ (c_{sl}^{\dagger}(\bm{r}))$ be an annihilation (creation) operator for an electron with spin $s(=\uparrow,\downarrow)$ at the position $\bm{r}=(x,y)$ of layer $l(=1,2)$.
The bilayer Rashba model (Fig.~\ref{fig:bilayer_system_schematic}) is given by the following Hamiltonian~\cite{yoshida2012pair}:
\begin{align}
    &\mathcal{H}
    =\frac{1}{2}\int d^2r_1d^2r_2\ \vec{C}^\dagger(\bm{r}_1)\check{H}_{\mathrm{BdG}}(\bm{r}_1,\bm{r}_2)\vec{C}(\bm{r}_2),\\
    &\check{H}_{\mathrm{BdG}}(\bm{r}_1,\bm{r}_2)
    =\delta(\bm{r}_1,\bm{r}_2)
    \check{H}^N(-i\nabla_2+e\bm{A}_2)+\check{\Delta}(\bm{r}_1,\bm{r}_2),\\
    \label{eq:mm-13}
    &\begin{aligned}
    \check{H}^N&(-i\nabla+e\bm{A})\\
    =&
    \left(\begin{array}{cc}
        \hat{H}^N(-i\nabla+e\bm{A})&\hat{0}\\
        \hat{0}&-\hat{H}^{N*}(-i\nabla+e\bm{A})
    \end{array}\right),
    \end{aligned}\\
    \label{eq:normal}
    &\begin{aligned}
    \hat{H}^N(\bm{k})
    =&\xi(\bm{k})\tilde{\sigma}_0\otimes\tilde{\tau}_0
    +\alpha\bm{g}(\bm{k})\cdot\tilde{\bm{\sigma}}\otimes\tilde{\tau}_z\\
    &
    +t_\perp\tilde{\sigma}_0\otimes\tilde{\tau}_x
    -\bm{h}\cdot\tilde{\bm{\sigma}}\otimes\tilde{\tau}_0,
    \end{aligned}\\
    \label{eq:mm-14}
    &\check{\Delta}(\bm{r}_1,\bm{r}_2)
    =\left(\begin{array}{cc}
        \hat{0}&\hat{\Delta}(\bm{r}_1,\bm{r}_2)\\
        -\hat{\Delta}^*(\bm{r}_1,\bm{r}_2)&\hat{0}
    \end{array}\right),
\end{align}
where $\vec{C}(\bm{r})=(\{c(\bm{r})\}, \{c^{\dagger}(\bm{r})\})^T$ with $\{c(\bm{r})\}=(c_{\uparrow 1}(\bm{r}),c_{\uparrow 2}(\bm{r}),c_{\downarrow 1}(\bm{r}),c_{\downarrow 2}(\bm{r}))^T$ and $\{c^{\dagger}(\bm{r})\}=(c^{\dagger}_{\uparrow 1}(\bm{r}),c^{\dagger}_{\uparrow 2}(\bm{r}),c^{\dagger}_{\downarrow 1}(\bm{r}),c^{\dagger}_{\downarrow 2}(\bm{r}))^T$.
We have introduced $\tilde{\sigma}_0,\tilde{\bm{\sigma}}=(\tilde{\sigma}_x,\tilde{\sigma}_y,\tilde{\sigma}_z)$ and $\tilde{\tau}_0,\tilde{\bm{\tau}}=(\tilde{\tau}_x,\tilde{\tau}_y,\tilde{\tau}_z)$ as the identity matrix and the Pauli matrices for spin and sublattice degrees of freedom, respectively.
In the normal Hamiltonian Eq.~\eqref{eq:normal}, $\xi(\bm{k})$ represents the two dimensional band dispersion in the layer, and $t_\perp$ denotes the inter-layer hopping.
The local inversion symmetry breaking in each layer is characterized by the staggered Rashba-type antisymmetric spin-orbit coupling (ASOC) $\bm{g}(\bm{k})=\bm{k}\times \hat{z}/k_{\mathrm{F}}$~\cite{yoshida2012pair,fischer2011superconductivity,yanase2022topological,sigrist2014superconductors,fischer2023superconductivity,maruyama2012locally} with $k_{\mathrm{F}}$ being the Fermi wave number.
The relative strength between $\alpha$ and $t_\perp$ serves as a parameter characterizing the system.
The magnetic field is applied perpendicular to the layers $\bm{H}=H\hat{\bm{z}}$ and the corresponding Landau gauge $\bm{A}(\bm{r})=Hx\hat{\bm{y}}$ is adopted.
The notation $\bm{h}=\mu_{\mathrm{B}}\bm{H}$ is used with the g-factor set to $g=2$.
Keeping in mind that the GL parameter $\kappa$ is estimated to be quite large $\gtrsim 10^2$ in $\mathrm{CeRh_2As_2}$~\cite{ogata2023parity,ogata2024appearance}, we consider the extreme type-II limit in which $\bm{B}=\bm{H}$ so that the screening effect due to the supercurrent is negligible.

The eigenvalues of the normal Hamiltonian $\hat{H}^N(\bm{k})$ are $E_{\nu,\lambda}(\bm{k})=\xi(\bm{k})+(-1)^{\nu}\sqrt{(t_\perp\pm \mu_{\mathrm{B}} H)^2+\alpha^2|\bm{g}(\bm{k})|^2}$, where $\nu=1,2$ is the band index and $\lambda=\pm$ is the index for pseudospin degrees of freedom.
In realistic cases, it is reasonable to assume $\mu_{\mathrm{B}}H,T_{\mathrm{c}0}\ll \alpha,t_\perp\ll E_{\mathrm{F}}$ for the energy scale of the system~\cite{lee2023linear,khim2021field,cavanagh2022nonsymmorphic}, where $T_{\mathrm{c}0}$ is the transition temperature at the zero magnetic field.
Then, when the Green's function is written in the band basis, the components with respect to different index $\nu$ can be ignored.
As a result, the effective low-energy Eilenberger equation~\cite{nagai2016multi} for almost doubly degenerate bands $\nu$ with $\lambda(=\pm)$ is given by
\begin{equation}
\label{eq:mm-6}
\begin{aligned}
    2\omega_n\tilde{f}^{(\nu)}
    +i(-1)^{\nu}h\cos\chi&[\tilde{\upsilon}_z,\tilde{f}^{(\nu)}]
    +i\bm{v}_{\mathrm{F}}\cdot\bm{\Pi}\tilde{f}^{(\nu)}\\
    =&\tilde{\Delta}^{(\nu)}\tilde{g}^{(\nu)}
    +\tilde{g}^{(\nu)}\tilde{\Delta}^{(\nu)},
\end{aligned}
\end{equation}
with the normalization condition $\tilde{g}^{(\nu)2}-\tilde{f}^{(\nu)}\tilde{f}^{(\nu)\circ}=\tilde{1}$.
Here, $\tilde{g}^{(\nu)}$ and $ \tilde{f}^{(\nu)}$ are normal and anomalous quasiclassical Green's functions for the band $\nu$, respectively.
In the Eilenberger equation, $\bm{v}_{\mathrm{F}}=\nabla_{\bm{k}}\xi$ is the Fermi velocity and we assume $\bm{v}_{\mathrm{F}}=v_{\mathrm{F}}\hat{\bm{k}}$ for the isotropic Fermi surface.
Also, $\bm{\Pi}=-i\nabla+2e\bm{A}$ is the gauge-invariant differential operator for the Cooper pair, $\omega_n=(2n+1)\pi T$ is the fermionic Matsubara frequency, $\tilde{\upsilon}_0,\tilde{\bm{\upsilon}}=(\tilde{\upsilon}_x,\tilde{\upsilon}_y,\tilde{\upsilon}_z)$ are the identity matrix and Pauli matrices for pseudospin degrees of freedom $\lambda$, and $[A,B]=AB-BA$.
In our formulation, the absolute values of $\alpha$ and $t_\perp$ themselves are not important and only their relative ratio appears through $\chi$, which is defined as $e^{i\chi}=(t_\perp+i\alpha\langle|\bm{g}|^2\rangle^{1/2})/\sqrt{t_\perp^2+\alpha^2\langle|\bm{g}|^2\rangle}$. This is because we assume those absolute values are sufficiently large compared to the energy scale of superconductivity.
The angle bracket $\langle\cdots\rangle$ denotes the angular average along the side of the cylindrical Fermi surface.
Note that the paramagnetic depairing effect appears as the effective Zeeman field $h_{\mathrm{eff}}\equiv h\cos\chi$ in Eq.~\eqref{eq:mm-6}.
The detailed derivation of Eq.~\eqref{eq:mm-6} is given in Appendix~\ref{sec:supp_1}.

We apply the "approximate solution"~\cite{adachi2005anisotropic,adachi2006basal,dan2015quasiclassical, adachi2005mixed}, an improvement on the Pesch approximation~\cite{pesch1975density}, to Eq.~\eqref{eq:mm-6}.
The approximate solution is based on the idea that the spatial variation of the normal Green's function is smaller than that of the anomalous Green's function~\cite{brandt1967theory}, and hence the action of $\bm{\Pi}$ on $\tilde{g}^{(\nu)}$ is neglected.
Then, Eq.~\eqref{eq:mm-6} is transformed into the following expression,
\begin{equation}
    \label{eq:mm-8}
    \tilde{f}^{(\nu)}
    \simeq\tilde{\Phi}^{(\nu)}\tilde{g}^{(\nu)}
    +\tilde{g}^{(\nu)}\tilde{\Phi}^{(\nu)},
\end{equation}
where $\tilde{\Phi}^{(\nu)}$ is defined as
\begin{equation}
    \label{eq:mm-7}
    \tilde{\Phi}^{(\nu)}
    =\{(2\omega_n+i\bm{v}_{\mathrm{F}}\cdot\bm{\Pi})\tilde{\upsilon}_0
    +i(-1)^{\nu}h\cos\chi[\tilde{\upsilon}_z,\cdot]\}^{-1}
    \tilde{\Delta}^{(\nu)}.
\end{equation}
For details on the "approximate solution", see Appendix~\ref{sec:supp_2} and \ref{sec:supp_6}.

In the following, the formation of $s$-wave intra-sublattice spin-singlet Cooper pairs is assumed.
The order parameters on the $l\,$th layer $\Delta_l$ are decomposed into the sublattice-symmetric component $\Delta_e=(\Delta_1+\Delta_2)/2$ and the antisymmetric component $\Delta_o=(\Delta_1-\Delta_2)/2$.
Then, $\Delta_j\ (j=e,o)$ are expanded in terms of Landau levels $\Psi_N$, which are the eigenfunctions of $\bm{\Pi}$,
\begin{align}
    \label{eq:mm-10}
    \Delta_j(\bm{k},\bm{r})
    =&w(\bm{k})\sum_{N}d_{j,N}\Psi_N(\bm{r})
    \quad(j=e,o),\\
    \label{eq:mm-11}
    \Psi_N(\bm{r})
    =&\sum_{m=-\infty}^{\infty}
    C_me^{iq_my}
    \psi_N(x/r_H+q_mr_H),\\
    \label{eq:mm-12}
    \psi_N(x)
    =&
    \frac{H_N(x)e^{-x^2/2}}{\sqrt{2^NN!}\pi^{1/4}}.
\end{align}
Here, $w(\bm{k})$ is the pairing function of the Cooper pair, $r_H=(2eH)^{-1/2}$ is the magnetic length, and $H_N$ is the $N$th Hermite polynomial.
This paper deals with $s$-wave superconductivity ($w(\bm{k})=1$), but we retain $w(\bm{k})$ in the equation below to demonstrate that other superconducting symmetries can be treated on an equal footing.
In Eq.~\eqref{eq:mm-10}, a triangular vortex lattice is assumed, since it is realized in most $s$-wave superconductors. This leads to $q_m=2\pi m/L_y$, where $L_y=2\sqrt{\pi}3^{-1/4}r_H$ is the period in $y$ direction and $C_m=(3\pi^2)^{1/8}e^{i\pi m^2/2}$ is the normalization constant.
Usually, in vortex states, the order parameter can be expressed as a superposition of low-order Landau levels, so only a few parameters need to be determined. 
The Landau level expansion~\eqref{eq:mm-10} yields an analytical expression of $\tilde{\Phi}^{(\nu)}$ for $\omega_n>0$ as
\begin{equation}
    \label{eq:mm-1}
    \begin{aligned}
    \tilde{\Phi}^{(\nu)}
    =&
    e^{-i\phi}w
    \int^{\infty}_{0}d\rho\ 
    e^{-2\omega_n\rho-|s|^2\rho^2/2}
    e^{-i(-1)^{\nu}\rho h\cos\chi[\tilde{\upsilon}_z,\cdot]}\\
    &\ \ \times
    \sum_{M,N}\Psi_M
    \mathcal{L}_{MN}(-is^*\rho)
    \tilde{d}^{(\nu)}_{N},
    \end{aligned}
\end{equation}
where
\begin{equation}
    \label{eq:mm-9}
    \mathcal{L}_{MN}(z)
    =
    \sum_{l=0}^{\mathrm{min}(M,N)}
    \frac{\sqrt{M!N!}}{(M-l)!(N-l)!l!}
    (z)^{M-l}(-z^*)^{N-l},
\end{equation}
$e^{i\phi(\bm{k})}=(g_x(\bm{k})+ig_y(\bm{k}))/|\bm{g}(\bm{k})|$ and $s=v_{\mathrm{F}}(\hat{k}_y-i\hat{k}_x)/(\sqrt{2}r_H)$.
The detailed derivation of Eq.~\eqref{eq:mm-1} is given in Appendix~\ref{sec:supp_2}.

To solve the Eilenberger equation self-consistently, we need gap equations for $\Delta_e$ and $\Delta_o$, projected onto each Landau level,
\begin{align}
    \label{eq:mm-2}
    &\frac{d_{e,N}}{V_e}
    =-\pi T
    \sum_{\nu}\frac{N_0}{2}
    \sum_{\omega_n>0}
    \left\langle w^*e^{i\phi}\mathrm{Tr}[\tilde{\upsilon}_x\overline{\Psi_N^*\tilde{f}^{(\nu)}}]\right\rangle,
    \\
    \label{eq:mm-3}
    &\begin{aligned}
    \frac{d_{o,N}}{V_o}
    =&-\pi T
    \sum_{\nu}\frac{N_0}{2}(-1)^{\nu}
    \sin\chi\\
    &\quad\quad\quad\quad\quad\quad\times
    \sum_{\omega_n>0}
    \left\langle w^*e^{i\phi}\mathrm{Tr}[\overline{\Psi_N^*\tilde{f}^{(\nu)}}]\right\rangle,
    \end{aligned}
\end{align}
where $V_{e}^{-1}=\sum_{\nu}(N_0/2)[\ln(T/T_{\mathrm{c}0}^{e})+2\pi T\sum_{\omega_n>0}\omega_{n}^{-1}]$ and $V_{o}^{-1}=\sin^2\chi \sum_{\nu}(N_0/2)[\ln(T/T_{\mathrm{c}0}^{o})+2\pi T\sum_{\omega_n>0}\omega_{n}^{-1}]$ are the strength of the attractive interaction in the symmetric channel for $\Delta_e$ and the antisymmetric channel for $\Delta_o$.
The overline in Eqs.~\eqref{eq:mm-2} and \eqref{eq:mm-3} denotes the spatial average, and $\mathrm{Tr}[\cdots]$ represents the trace of the matrix.
$T_{\mathrm{c}0}^{e}$ and $T_{\mathrm{c}0}^{o}$ stand for the transition temperature of the $\Delta_e$ and $\Delta_o$ channel at zero magnetic field, respectively.
The density of states in the normal state for bands $\nu=1,2$ are assumed to be equal: $N_{0}^{(1)}=N_{0}^{(2)}=N_0$.
The detailed derivation of Eqs.~\eqref{eq:mm-2} and \eqref{eq:mm-3} is given in Appendix~\ref{sec:supp_3}.

The numerical procedure for solving the quasiclassical equation self-consistently at each $(H,T)$ is as follows~\cite{adachi2006basal,adachi2005anisotropic}:
(i) give the initial value of $\{d_{j,N}\}$, 
(ii) solve the simultaneous equations of Eq.~\eqref{eq:mm-8} and the normalization condition for each $\bm{k}_{\mathrm{F}},\bm{r},i\omega_n$ to obtain $\tilde{g}^{(\nu)}$ and $\tilde{f}^{(\nu)}$, 
(iii) update $\{d_{j,N}\}$ by using Eqs.~\eqref{eq:mm-2} and \eqref{eq:mm-3}, 
(iv) return to step (ii) and repeat these steps until self-consistency is achieved.
Note that when either $\Delta_e$ or $\Delta_o$ is zero, step (ii) can be performed analytically (see Appendix~\ref{sec:supp_2} for details).

Various physical quantities can be computed using the self-consistent solution $\tilde{g}^{(\nu)}_{\mathrm{sc}}$, $\tilde{f}^{(\nu)}_{\mathrm{sc}}$ and $\{d_{j,N,\mathrm{sc}}\}$ of these equations.
In the following, the order parameter obtained by the self-consistent solution via Eq.~\eqref{eq:mm-10} is denoted by $\Delta_{j,\mathrm{sc}}$.

First, the superconducting free energy relative to the normal state is calculated by
\begin{widetext}
\begin{equation}
\label{eq:mm-4}
\begin{aligned}
    \delta\mathcal{F}
    =&\mathcal{F}_S-\mathcal{F}_N\\
    =&4\Omega
    \sum_{\nu}\frac{N_0}{2}\sum_{N}
    \left[
    \frac{|d_{e,N}|^2}{V_eN_0}
    +2\pi T\sum_{\omega_n>0}
    \mathrm{Re}\int^{1}_{0}dx_e\ d_{e,N}^{*}
    \langle e^{i\phi}w^*
    \mathrm{Tr}[\tilde{\upsilon}_x\overline{\Psi_N^*\tilde{f}^{(\nu)}(x_e)}]\rangle
    \right.\\
    &\quad\quad\quad\quad+
    \left.
    \frac{|d_{o,N}|^2}{V_oN_0}
    +2\pi T\sum_{\omega_n>0}
    (-1)^{\nu}\mathrm{Re}\int^{1}_{0}dx_o\ d_{o,N}^{*}
    \langle e^{i\phi}w^*\sin\chi
    \mathrm{Tr}[\overline{\Psi_N^*\tilde{f}^{(\nu)}(x_o)}]\rangle
    \right],
\end{aligned}
\end{equation}
\end{widetext}
which is required to investigate the phase diagram.
Here, $\Omega$ is the area of each layer, and $\tilde{f}^{(\nu)}(x_j)$ is the non-self-consistent solution of Eq.~\eqref{eq:mm-8} obtained by replacing $\Delta_j$ with $x_j\Delta_{j,\mathrm{sc}}$.
The detailed derivation of Eq.~\eqref{eq:mm-4} is given in Appendix~\ref{sec:supp_4}.
Assuming a second-order superconducting phase transition, $\delta\mathcal{F}$ can be expanded with respect to the order parameter near the upper critical field, and its quadratic term $\delta\mathcal{F}^{(2)}$ is given in the form of
\begin{equation}
\label{eq:mm-5}
    \delta\mathcal{F}^{(2)}
    =4\Omega\sum_{\nu}\frac{N_0}{2}
    \sum_{j,N}
    |d_{j,N}|^2E_{j,N},
\end{equation}
where $E_{j,N}$ are the excitation energies of the modes corresponding to $d_{j,N}$.
The derivation of Eq.~\eqref{eq:mm-5} and the explicit expression for $E_{j,N}$ are given in Appendix~\ref{sec:supp_4}.
Note that Eq.~\eqref{eq:mm-5} is identical to the result of the previous study~\cite{minamide2025superconducting}.
Then, the Maki parameter~\cite{maki1966effect} is represented as $\alpha_{\mathrm{M}}=2\sqrt{2}\pi T_{\mathrm{c}0}^{e}\cos\chi/(mv_{\mathrm{F}}^2)$ in the bilayer Rashba model, where $m$ is the electron mass.
Since Eq.~\eqref{eq:mm-5} is diagonal with respect to the Landau level index $N$, the upper critical field $H_{c2}^j(T)$ can be obtained by solving equation $E_{j,N=0}(H,T)=0$, given that the lowest Landau level with $N=0$ orders at the upper critical field.

Next, we also calculate the local density of states (LDOS) as it is experimentally observable and indicative of the local properties of vortex states.
The retarded quasiclassical Green's function $\hat{g}^{(\nu)R}(\bm{k}_{\mathrm{F}},\bm{r},E)$ is obtained by solving the simultaneous equations of Eq.~\eqref{eq:mm-8} and the normalization condition with analytic continuation $i\omega_n\to E+i\eta$, where $\eta>0$ is the smearing parameter.
The order parameter is given by the substitution $d_{j,N}=d_{j,N,\mathrm{sc}}$.
Then, we reach the expression of the LDOS,
\begin{equation}
    \label{eq:mm-15}
    N(\bm{r},E)
    =
    \sum_{\nu}\frac{N_0}{2}
    \mathrm{Re}\,\langle \mathrm{Tr}\,\tilde{g}^{(\nu)R}(\bm{k}_{\mathrm{F}},\bm{r},E)\rangle.
\end{equation}
The derivation of Eq.~\eqref{eq:mm-15} is given in Appendix~\ref{sec:supp_5}.

The details of the numerical calculations are as follows. 
The Broyden's method is used in the self-consistent calculation.
Also, the Bohr magneton is set to $\mu_{\mathrm{B}}=1/2$.
The quasiclassical Green's functions are evaluated at $41\times 41$ discretized points in a unit cell of the vortex lattice, and the cut-off energy regarding the Matsubara frequency is set to $\varepsilon_{\rm c}=40\,T_{\mathrm{c}0}^{e}$.
Contributions from Landau levels up to the $12$-th order are taken into account, unless otherwise stated.
We have examined how truncating the Landau levels and Matsubara frequencies to finite values affects the numerical results in Appendix~\ref{sec:supp_6}.

\section{Results and Discussion}
\label{sec:results_and_discussion}

Our quasiclassical equations have three material parameters: the relative strength of the ASOC and the interlayer hopping $\alpha/t_\perp$ [or $\chi=\tan^{-1}(\alpha/t_\perp)$], the Maki parameter $\alpha_{\mathrm{M}}$, and the ratio of transition temperatures at zero magnetic field for the BCS and PDW states $T_{\mathrm{c}0}^{o}/T_{\mathrm{c}0}^{e}$.
In what follows, we set these parameters to $\alpha/t_\perp=15.0,\ \alpha_{\mathrm{M}}=4.8,\ T_{\mathrm{c}0}^{o}/T_{\mathrm{c}0}^{e}=0.9$, the same values as in the previous study~\cite{minamide2025superconducting}.

\subsection{Phase diagram}
\begin{figure}[tbp]
  \centering
\includegraphics[width=0.48\textwidth]{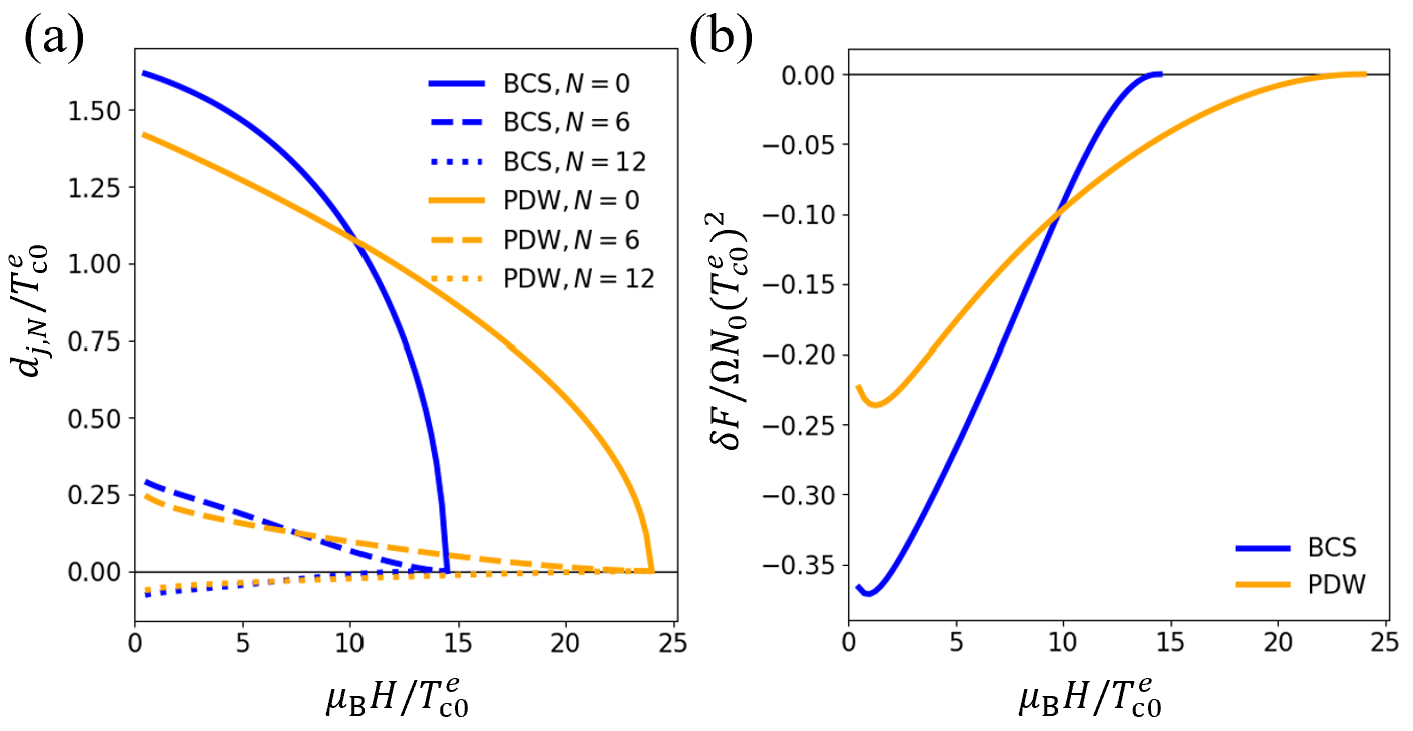}
  \caption{The magnetic field-dependence of (a) order parameters and (b) free energy of superconducting states at $T/T_{\mathrm{c}0}^{e}=0.5$.
  In (a), the blue and orange lines show the order parameters $\{d_{j,N}\}$ for the BCS ($j=e$) and PDW ($j=o$) states, respectively.
  The solid, dashed, and dotted lines represents the order parameters for the Landau level indices $N=0,\,6,\,12$, respectively.
  In (b), the free energies of the BCS and PDW states are shown in the blue and orange lines, respectively.
  The energies are measured in the unit of $\Omega N_0 (T_{\mathrm{c}0}^{e})^2$.}
\label{fig:op_and_ce}
\end{figure}

The mean-field theory~\cite{yoshida2012pair} for the spatially uniform states has elucidated the phase diagram of locally noncentrosymmetric layered superconductors in the Pauli limit. The BCS state with sublattice-symmetric spin-singlet order parameter $[(\Delta_1,\Delta_2)=(\Delta,\Delta)\Leftrightarrow (\Delta_e,\Delta_o)=(\Delta,0)]$ and the PDW state with sublattice-antisymmetric spin-singlet order parameter $[(\Delta_1,\Delta_2)=(\Delta,-\Delta)\Leftrightarrow (\Delta_e,\Delta_o)=(0,\Delta)]$ are stabilized, respectively, on the low- and high-field side of the phase transition line in the superconducting state.
In this subsection, the phase diagram of the bilayer Rashba model is determined with the orbital depairing effect by using the quasiclassical theory introduced in the previous section.
Figure~\ref{fig:op_and_ce} shows the magnetic field-dependence of the order parameters and free energy gains for the BCS and PDW states at $T/T_{\mathrm{c}0}^{e}=0.5$.
In Fig.~\ref{fig:op_and_ce}(a), the blue (orange) solid, dashed, and dotted curves represent the $0$-th, $6$-th and $12$-th components of the order parameter $\{d_{j,N}\}$ for the BCS (PDW) state, respectively.
Here, the coefficients $\{d_{j,N}\}$ are taken to be real by using the $U(1)$ gauge symmetry.
Since a triangular vortex lattice is assumed, only the components of multiple orders of $6$ are relevant~\cite{adachi2006basal,watanabe2005magnetic}.
As expected, the amplitude of order parameters gradually develops as the magnetic field decreases.
The $0$-th component, the order parameter corresponding to the lowest Landau level, grows with an upward convex curve, while the $6$-th and $12$-th components grow linearly or with a slightly downward convex curve.
Consequently, the ratios of the higher Landau level components $d_{j,6}/d_{j,0},\,d_{j,12}/d_{j,0}$ increase toward the low field region.
In Fig.~\ref{fig:op_and_ce}(b), the blue (orange) curve indicates the free energy of the BCS (PDW) state relative to the normal state.
The unphysical upward curvatures seen below $\mu_{\mathrm{B}} H/T_{\mathrm{c}0}^{e}\simeq 1.0$ for both BCS and PDW states imply the collapse of the picture of the Landau quantization around the zero field.
We will not address this issue below because this study focuses on the region with much higher magnetic fields.
The point where the relative free energy vanishes corresponds to the normal-superconducting phase transition.

\begin{figure}[tbp]
  \centering
  \includegraphics[width=0.42\textwidth]{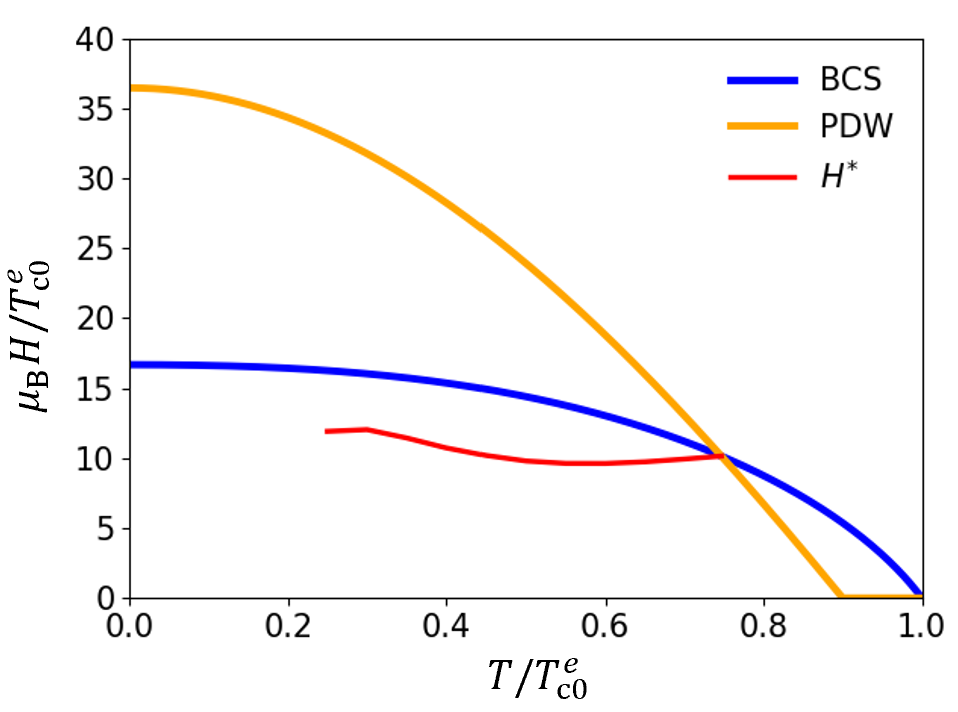}
  \caption{The phase diagram of the bilayer Rashba model with $\alpha/t_\perp=15.0,\,\alpha_{\mathrm{M}}=4.8$ and $T_{\mathrm{c}0}^{o}/T_{\mathrm{c}0}^{e}=0.9$.
  The red line represents the parity transition line $H^*(T)$ down to $T/T_{\mathrm{c}0}^{e}=0.25$.
  The blue and orange lines indicate the upper critical fields of the BCS and PDW states, respectively.
  }
\label{fig:phase_diagram}
\end{figure}

Because the Maki parameter $\alpha_{\mathrm{M}}$ is large enough~\cite{minamide2025superconducting}, the results are consistent with the Pauli-limit picture: in Fig.~\ref{fig:op_and_ce}(b), the BCS state has lower free energy at low magnetic field region, whereas the PDW state survives to high field regimes.
Thus, the parity transition is supported by the quasiclassical theory for the vortex lattice states.
Performing the same calculations as for Fig.~\ref{fig:op_and_ce} at each temperature, we obtain the phase diagram shown in Fig.~\ref{fig:phase_diagram}. 
The upper critical fields $H_{c2}^{j}(T)$ of the BCS and PDW states are drawn with the blue and orange lines, respectively.
Their intersection gives rise to a kink in the experimentally observable upper critical field of CeRh$_2$As$_2$~\cite{khim2021field}.
As noted in the previous section, since the expression of $E_{j,N}$ is identical to that in the GL theory, the upper critical fields, obtained from the condition $E_{j,N=0}=0$, are exactly the same as the result in the previous study~\cite{minamide2025superconducting}.
Meanwhile, the crossing point of the free-energy curves defines the parity transition field $H^*$.
By finding $H^*$ at each temperature, the parity transition line is determined down to the low temperature region, where the GL theory~\cite{minamide2025superconducting} could not reach.
In Fig.~\ref{fig:phase_diagram}, the red line shows the parity transition line $H^*(T)$.
The parity transition line is not monotonic: it shows positive and negative slopes in the high and low temperature regions, respectively.
This feature is not compatible with the internal phase transition line in the experimentally obtained phase diagram of $\mathrm{CeRh_2As_2}$, which is almost horizontal~\cite{khim2021field}, but has a slightly positive slope~\cite{khanenko2025phase}.
Although the discrepancy is not significant, it could result from a number of factors that are not incorporated into our simplified model, such as the anisotropy of the Fermi surface and the superconducting pairing symmetry, and these effects will be discussed further in Sec.~\ref{sec:conclusion}.
In Fig.~\ref{fig:phase_diagram}, the parity transition line is not plotted below $T/T_{\mathrm{c}0}^{e}=0.25$, due to the poor convergence of the self-consistent calculation.
This issue originates from the fact that the "approximate solution" is not applicable in the low-temperature region far below the upper critical field (see Appendix~\ref{sec:supp_6} for a detailed discussion regarding the reliability of the approximation).
The quasiclassical theory itself remains valid at low temperatures. Hence, the parity transition line can, in principle, be calculated at even lower temperatures by resorting to the full solution.
However, since the Maki parameter is significantly larger than unity, it is expected that the transition line will be approximately horizontal, albeit with some slope, similar to the results obtained in the Pauli limit~\cite{yoshida2012pair}.

\subsection{Local properties of vortex states}
In this subsection, we show the spatially resolved characteristics of the vortex states and clarify the differences between the BCS and PDW superconducting states. 
Furthermore, we characterize the superconducting meron state from the perspective of vortex bound states.

\begin{figure}[tbp]
  \centering
  \includegraphics[width=0.375\textwidth]{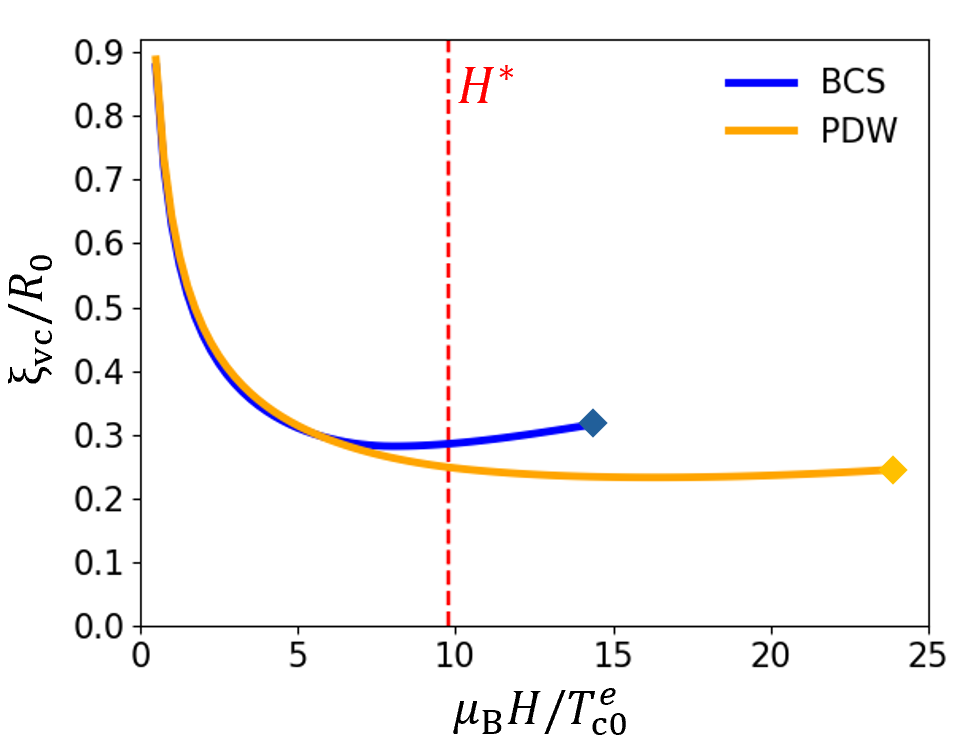}
  \caption{The magnetic field-dependence of the vortex core radii for the BCS (blue) and PDW (orange) states.
  The length is measured in the unit of $R_0=v_{\mathrm{F}}/T_{\mathrm{c}0}^{e}$.
  The temperature is fixed at $T/T_{\rm c0}^{e}=0.5$ and the minimum value of the magnetic field is $\mu_{\mathrm{B}} H/T_{\mathrm{c}0}^{e}=0.5$.
  The endpoints of the curves on the high-field side, indicated by diamonds, correspond to the upper critical fields $H_{c2}^{j}$.
  The red vertical line represents the parity transition field $H^*$.
  }
\label{fig:vc_radius}
\end{figure}

First, we discuss the vortex core radius.
The definition of the vortex core radius $\xi_{\mathrm{vc}}$ varies in the literature, but here we estimate it from the initial slope of the pair potential at the vortex core~\cite{ichioka1999vortex,higashi2016robust,sonier2000musr,hayashi2005impurity,kramer1974core} as
\begin{equation}
\label{eq:res-3}
    \xi_{\mathrm{vc}}=\Delta_{\mathrm{max}}/\lim_{|\bm{r}|\to 0}\frac{|\Delta(\bm{r}+\bm{r}_{\mathrm{vc}})|}{|\bm{r}|},
\end{equation}
where $\bm{r}_{\mathrm{vc}}$ is the position of a vortex and $\Delta_{\mathrm{max}}$ is the amplitude of the pair potential at the midpoint of the nearest-neighbor vortices.
The anisotropy of the definition in Eq.~\eqref{eq:res-3} due to the vortex lattice effect need not be considered since the pair potential is isotropic sufficiently close to a vortex core.
The magnetic field-dependence of the vortex core radius $\xi_{\mathrm{vc}}(H)$, normalized by $R_0\equiv v_{\mathrm{F}}/T_{\mathrm{c}0}^{e}$ for the BCS and PDW states, is shown in Fig.~\ref{fig:vc_radius}.
With increasing magnetic field, the vortex core radius gradually decreases because the vortices approach each other, approximately following $r_H\propto H^{-1/2}$, which is typical for many superconductors~\cite{ichioka1999vortex,ichioka1999field,kogan2005field,sonier1997muon,fente2016field,callaghan2005field,kogan2006effect}.
However, in the BCS state, the vortex core radius reaches a minimum at an intermediate magnetic field ($\mu_{\mathrm{B}} H/T_{\mathrm{c}0}^{e}\simeq 8.0$ for $T/T_{\mathrm{c}0}^{e}=0.5$), above which it begins to grow again.
The origin of this phenomenon can be ascribed to the paramagnetic depairing effect as pointed out by Higashi \textit{et al.}~\cite{higashi2016robust}: the paramagnetic depairing effect causes the energy dispersion to be particle-hole asymmetric in each mirror subsector of the BdG Hamiltonian, effectively reducing the size of the superconducting gap.
Then, the coherence length, at which the pair potential restores from the vortex core, increases because it is inversely proportional to the superconducting gap.
Therefore, the competition between the orbital and paramagnetic depairing effects gives rise to the non-monotonic magnetic field dependence of the vortex core radius in the BCS state.
This interpretation is consistent with the absence of a notable increase in the vortex core radius in the PDW state because the paramagnetic depairing effect is almost negligible.
These different behaviors of the BCS and PDW states result in a sudden shrinkage of the vortex core radius at the superconducting parity transition. 
The percentage of the shrinkage $\delta_{\mathrm{vc}}(T)\equiv 1-\xi_{\mathrm{vc, PDW}}/\xi_{\mathrm{vc, BCS}}$, where the right hand side is evaluated at $H=H^*(T)$, increases with lowering the temperature and reaches $45\,\%$ at $T/T_{\mathrm{c}0}^{e}=0.25$.
Previous studies~\cite{mockli2018orbitally,higashi2016robust} have made contradictory predictions about the change in core size associated with the parity transition, but we believe that our result settles this controversy because our formulation resolves unverified issues of the previous theories: in this study, the quasiclassical theory is formulated on the basis of a microscopic model without relying on phenomenological parameters, and it takes into account the influence of the vortex lattice.

\begin{figure*}[tbp]
  \centering
  \includegraphics[width=0.75\textwidth]{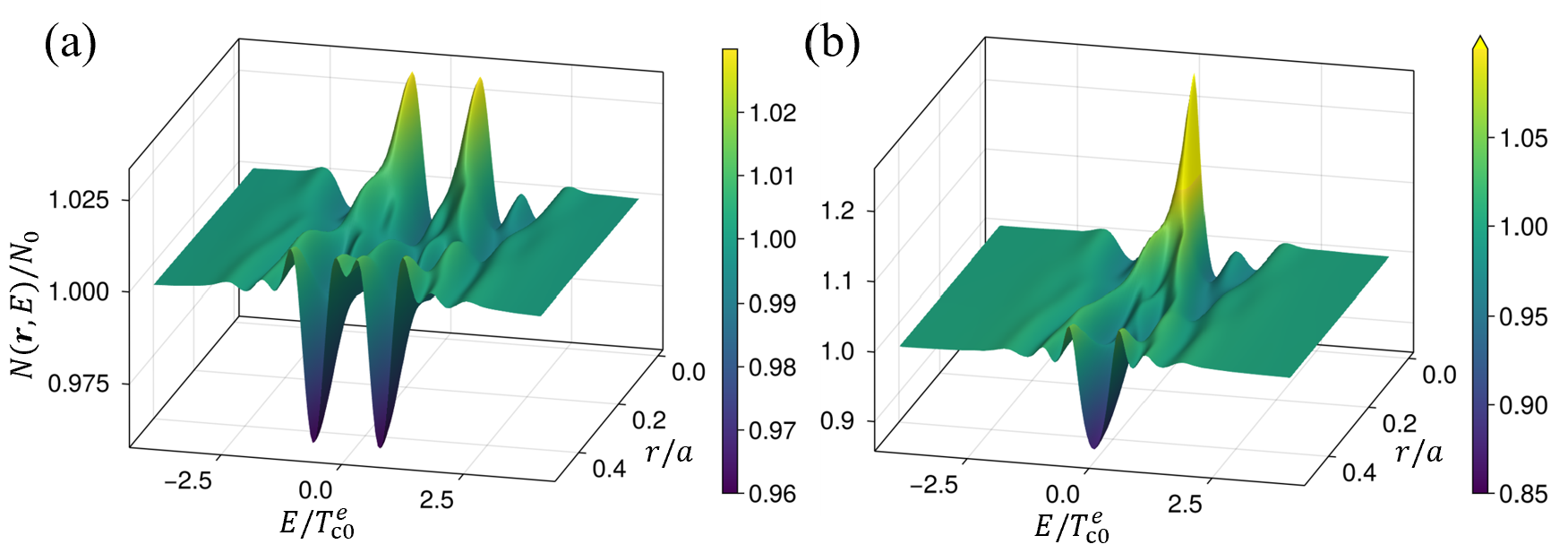}
  \caption{The LDOS near the vortex core for the (a) BCS and (b) PDW states.
  The back and front sides of the figure correspond to the vortex core and the midpoint of the nearest-neighboring vortices, respectively.
  The distance from the vortex core $r$ is measured in the unit of $a$, which is the distance between nearest-neighboring vortices.
  The temperature and magnetic field are fixed at $T/T_{\mathrm{c}0}^{e}=0.74,\ \mu_{\mathrm{B}} H/T_{\mathrm{c}0}^{e}=10.25$, which is located near the multicritical point in the $H$-$T$ phase diagram (Fig.~\ref{fig:phase_diagram}).
  }
\label{fig:ldos_bcs_pdw}
\end{figure*}

Next, the quasiparticle bound states around a vortex core are studied. 
In regions away from the upper critical field, the approximate solution overestimates the magnitude of the order parameter compared to the exact solution~\cite{adachi2006basal}.
This is not a major problem as far as thermodynamic quantities are concerned.
However, it leads to less precise results when calculating fine spatial structures, because the exact values of the amplitudes of higher Landau levels are crucial.
Moreover, the previous study~\cite{dan2015quasiclassical} showed that in such a region, a large smearing parameter is required to obtain the physical solution although the clean limit is considered.
For these reasons, we focus on a temperature and a magnetic field near the upper critical field to calculate the LDOS, where the amplitude of higher Landau levels is quite small and the "approximate solution" is almost exact.
In the following, we typically use the smearing parameter $\eta=0.03\,T_{\mathrm{c}0}^{e}$.
Figure~\ref{fig:ldos_bcs_pdw} shows the LDOS of the BCS and PDW states.
The vertical and horizontal axes represent the LDOS and the quasiparticle energy, respectively.
The depth direction represents the distance from the vortex core, with the back and front sides of the figure corresponding to the position at a vortex core and the midpoint of nearest-neighboring vortices, respectively.
The temperature and magnetic field are fixed at $T/T_{\mathrm{c}0}^{e}=0.74,\ \mu_{\mathrm{B}} H/T_{\mathrm{c}0}^{e}=10.25$, which is close to the multicritical point located at $T/T_{\mathrm{c}0}^{e}=0.747,\ \mu_{\mathrm{B}} H/T_{\mathrm{c}0}^{e}=10.14$.
In both BCS and PDW states, the LDOS has a substantial spatially uniform component, since the magnetic field is set to be close to the upper critical field~\cite{brandt1967theory,ichioka1999field}.
The difference between the two superconducting states is evident in the peak structure of the LDOS at the vortex core.
In the BCS state, there are two peaks that are split up and down from zero energy due to the paramagnetic depairing effect~\cite{ichioka2007vortex_FFLO,mizushima2005topological,ichioka2007vortex_Pauli,higashi2016robust}.
By calculating the spin-resolved LDOS, we find that the peak positions are estimated to be $E/T_{\mathrm{c}0}^{e}\simeq\pm 0.70$ for up and down spins, respectively. These energies agree with the effective Zeeman field $h_{\mathrm{eff}}=h\cos\chi\simeq 0.68\,T_{\mathrm{c}0}^{e}$.
For the derivation of the spin-resolved LDOS and its numerical results, see Appendix~\ref{sec:supp_5}.
The PDW state, on the other hand, exhibits a large peak of LDOS at zero energy, because the equal pseudospin pairing formation completely suppresses the paramagnetic depairing effect~\cite{yoshida2012pair}.
As one moves away from the vortex core, this zero-energy peak gradually splits.
These qualitative differences in the spatial dependence of LDOS between the BCS and PDW states are expected to be universal and robust, as it originates from the difference in the intrinsic properties of Cooper pairs in each superconducting state, which determines whether the paramagnetic depairing effect is effective.
In particular, the resulting drastic change in the LDOS at the parity transition should be observable by experimental probes sensitive to local quasiparticle spectra, as discussed later.
Finally, we comment on the relation to the previous study dealing with the single-vortex system~\cite{higashi2016robust}.
Our results are qualitatively consistent with this study, but there are two differences.
First, in our formulation, the Zeeman field is renormalized by the SOC as $h\to h_{\mathrm{eff}}$.
Second, while four distinct peaks appear at the vortex core in the BCS state in the result of Higashi \textit{et al.} [Fig.~3(a) in \cite{higashi2016robust}], the outer two peaks are pushed to high energies as the SOC is increased. 
In the strong-SOC limit considered in this paper, these outer peaks lie outside the low-energy range of interest, and only the inner two peaks remain visible, which is consistent with our results.

\begin{figure*}[tbp]
  \centering
  \includegraphics[width=0.7\textwidth]{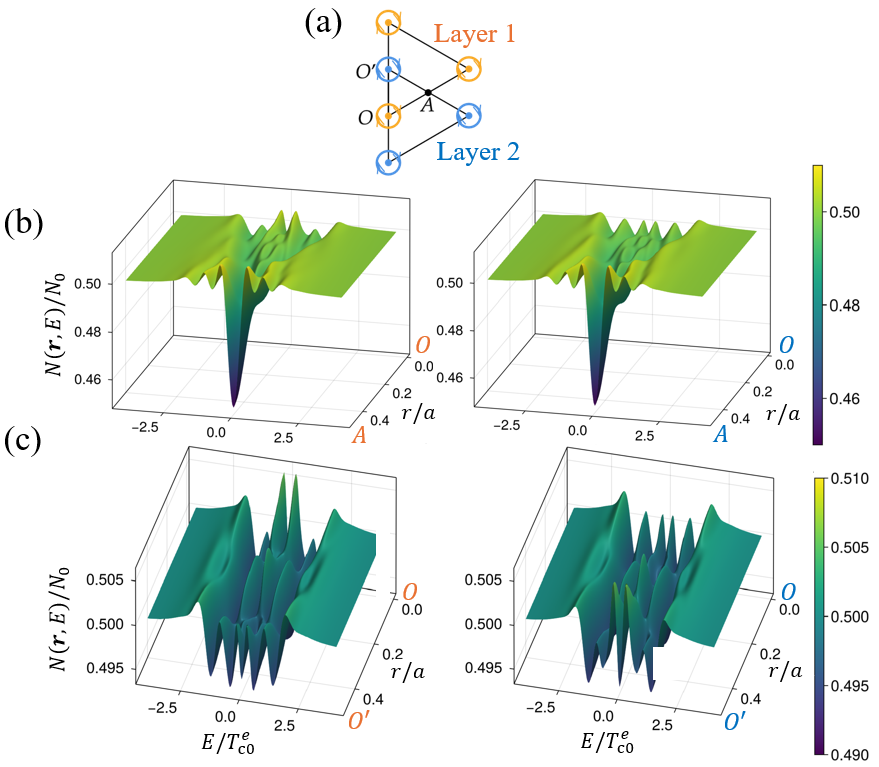}
  \caption{(a) The characteristic positions $O$, $O'$, and $A$ in the vortex lattice of the SCM state viewed from a direction perpendicular to the layer.
  The orange and blue circles represent the position of vortices on the layer $1$ and $2$, respectively.
  The layer-resolved LDOS for the SCM state along the lines $OA$ and $OO'$ is shown in (b) and (c), respectively.
  The left (right) panels show the LDOS in the layer $1$ ($2$).
  The temperature and magnetic field are fixed at $T/T_{\mathrm{c}0}^{e}=0.74,\ \mu_{\mathrm{B}} H/T_{\mathrm{c}0}^{e}=10.25$, as in Fig.~\ref{fig:ldos_bcs_pdw}.
  Note that non-self-consistent order parameters are adopted in this calculation.
  }
\label{fig:ldos_meron}
\end{figure*}

So far, we focused on the BCS and PDW vortex states which are stable in locally noncentrosymmetric superconductors in a wide range of parameters.
In the next part of this subsection, we discuss a novel superconducting state that we have named the superconducting meron (SCM) state~\cite{minamide2025superconducting}.
The order parameter of this state is given by 
\begin{equation}
    \label{eq:res-4}
    \begin{aligned}
    \Delta_l(\bm{r})&\propto
    \sum_{m=-\infty}^{\infty}
    \left[e^{iq_{2m}y}\psi_0(x/r_H+q_{2m}r_H)\right.\\
    &\left.+(-1)^{l+1}i\,e^{iq_{2m+1}y}\psi_0(x/r_H+q_{2m+1}r_H)
    \right],
    \end{aligned}
\end{equation}
which has both sublattice-symmetric $\Delta_e$ and antisymmetric $\Delta_o$ components, while the BCS and PDW states have only either of them.
In this sense, the SCM state is a multi-component superconducting state that has no counterpart in the uniform superconducting state because the even and odd parity components coexist in a spatially inhomogeneous manner. 
The spatial structure is characterized by the layer-dependent vortex position.
In each layer, the vortex cores are arranged in a triangular lattice as in the BCS and PDW states, but their positions are shifted horizontally by half of the primitive lattice vector with respect to the adjacent layer [see Fig.~\ref{fig:ldos_meron}(a)].
Across adjacent layers, the magnetic flux lines are distorted so as to link vortices in neighboring layers, meaning that these vortices share a continuous flux line.
The magnetic flux threading each vortex within an individual layer remains equal to a single flux quantum.
The three-component unit vector $\bm{n}(\bm{r})=\chi^{\dagger}\bm{\sigma}\chi/\chi^{\dagger}\chi$~\cite{dimitrova2007theory,garaud2015properties,zhang2016electronic,fernandez2016vortical,babaev2002hidden,mueller2004spin,kasamatsu2004vortex,kasamatsu2005spin,tsubota2010quantized,hu2015half,zyuzin2017nematic,noda2025fractional,garaud2013chiral} is defined from the two-component layer-dependent order parameter $\chi(\bm{r})=(\Delta_1(\bm{r}),\Delta_2(\bm{r}))^{T}$ at each two-dimensional position on the layers, and then the configuration of $\bm{n}$ is characterized by the "meron"-lattice pattern, where the meron refers to the topological spin texture with one half of the Skyrmion number~\cite{kosevich1990magnetic,nagaosa2013topological,finocchio2016magnetic,gobel2021beyond}.

According to the microscopically derived GL theory~\cite{minamide2025superconducting}, the SCM state is stable in a region near the multicritical point.
However, in the quasiclassical calculation used in this study, since the number of equations is larger than that of the independent variables due to the constraints $|\Delta_e|=|\Delta_o|$, a self-consistent solution could not be obtained for the SCM state.
Therefore, instead, the LDOS in the SCM state is calculated by assuming reasonable order parameters that are estimated to be comparable to those for the BCS and PDW states.
For simplicity, we assume that only the lowest Landau levels are included in the order parameters and their magnitudes are $d_{e,0}=d_{o,0}=0.15$.
Because the free energies of the SCM state and the other two competing states are anticipated to be of similar magnitude in this parameter region~\cite{minamide2025superconducting}, it is reasonable to assume that the dominant lowest-Landau-level component has a similar magnitude in all three states.
Although this order parameter is not obtained self-consistently, we have confirmed that the qualitative results remain unchanged unless the order parameters are significantly changed.
Figure~\ref{fig:ldos_meron}(b) shows the layer-resolved LDOS along the line $OA$ illustrated in Fig.~\ref{fig:ldos_meron}(a) (see Appendix~\ref{sec:supp_5} for the derivation of the layer-resolved LDOS).
In the layer $1$ (left panel), the LDOS exhibits a peak structure in the low-energy region at the vortex core, although it is less sharp than the PDW and BCS states.
This can be understood because the vortex core in one layer coincides with a position in the neighboring layer that is farthest from its vortex core, where the superconducting gap attains its maximum value. 
Therefore, the interlayer coupling at such positions is expected to diminish the zero-bias peak.
In addition, the paramagnetic depairing effect can suppress the zero-bias peak, as in the BCS state.
What is interesting is that even though the vortex core is not located at the point $O$ in the layer $2$, there are weak but visible peaks in the LDOS [right panel of Fig.~\ref{fig:ldos_meron}(b)].
The situation is better illustrated in Fig.~\ref{fig:ldos_meron}(c), which displays the layer-resolved LDOS along the line $OO'$.
We see a characteristic behavior: a ridge-shaped structure connecting the vortex cores in layer 1 ($O$) and layer 2 ($O'$) appears in the low-energy LDOS.
This indicates significant quasiparticle transfer between $O$ and $O'$ by the interlayer coupling $t_\perp$, since the inter-vortex distance is the smallest in this horizontal direction.

\section{Discussion and Conclusion}
\label{sec:conclusion}
As we have discussed, the multiple superconducting phases in $\mathrm{CeRh_2As_2}$ are well understood by the parity transition scenario, in which the low-field and high-field phases correspond to the BCS and PDW states, respectively.
However, in fact, direct experimental evidence to identify the order parameter of the superconducting states has not yet been obtained.
Furthermore, no signature of the theoretically predicted SCM state has been found, and this may be because the SCM state is stabilized only in a narrow magnetic field region~\cite{minamide2025superconducting}.
This current situation calls for an experimental probe to clarify the superconducting states in $\mathrm{CeRh_2As_2}$ in more detail.
The magnetic susceptibility was proposed in an early study~\cite{maruyama2012locally}: the magnetic susceptibility in the BCS state decreases with temperature as it does in the ordinary spin-singlet superconducting state, whereas in the PDW state it remains unchanged from the normal state.
Although the subsequent experiment~\cite{ogata2023parity} did not show the behaviors theoretically predicted, the recent experiment~\cite{ogataprivate} using high-quality samples supported the prediction in the zero-field BCS and high-field PDW states. However, the experiment did not detect the discontinuous change at the parity transition, possibly due to the antiferromagnetic order~\cite{ogata2023parity,ogata2024appearance,kibune2022observation,khim2025coexistence,chajewski2024discovery} and the potential quadrupole order~\cite{hafner2022possible} that coexist with superconductivity in CeRh$_2$As$_2$ at ambient pressure. Thus, the exploration of other quantities related to superconducting parity transition has been awaited.

The measurement of quasiparticle bound states near vortex cores has a long history of use in identifying the structure of order parameters.
Therefore, it could serve as an experimental probe to verify the superconducting parity transition and find the SCM state.
Scanning tunneling microscopy-spectroscopy (STM-STS), small-angle neutron scattering (SANS), nuclear magnetic resonance (NMR), and muon spin relaxation ($\mu$SR) measurements are potential methods for accessing the LDOS, the vortex core radius, and the vortex lattice structure.
In particular, the presence or absence of the paramagnetic depairing effect on the LDOS can be directly observed via spin-polarized STM~\cite{wiesendanger2009spin,choi2017switching,wang2021spin}.
Furthermore, the flux flow resistivity measurement can detect the superconducting parity transition and the meron state, since the flux flow reflects the pairing symmetry, the Fermi surface anisotropy, and information on vortex bound states~\cite{blatter1994vortices,golosovsky1996high,takaki2002effects,kato2002numerical,kopnin2001theory,kato2000phase,kopnin1997flux,higashi2013field,bardeen1965theory}.

In conclusion, we investigated the vortex states of locally noncentrosymemtric superconductors by employing the quasiclassical theory.
The phase diagram with multiple superconducting phases was obtained, and in particular, the superconducting parity transition was clarified, including the orbital effect due to the gauge field.
Then, the spatially-resolved properties of vortices in each superconducting phase were examined, and two noteworthy phenomena were revealed to characterize them: the sudden shrinkage of vortex core radius at the parity transition and the vortex bound states reflecting the microscopic pairing symmetry.
These results open a way to explore the nature of multiple superconducting phases in $\mathrm{CeRh_2As_2}$.

Finally, we address three features of $\mathrm{CeRh_2As_2}$ that are not accounted for in this paper and provide future prospects.
(1) This study ignored all the anisotropy of the pair potential and the Fermi surface. 
In fact, $\mathrm{CeRh_2As_2}$ has a complex Fermi surface consisting of several bands, and theoretical works predict $d$-wave superconductivity~\cite{nogaki2022even,lee2025unified,nogaki2024field}.
It is therefore important to clarify how material anisotropies affect the physical quantities discussed in this work.
(1-a) The anisotropies lead to Landau-level mixing in the quadratic term of the free energy with respect to the order parameter. 
As a result, the upper critical field $H_{c2}$ is no longer determined solely by the lowest Landau level, and its value tends to be enhanced at low temperatures~\cite{dan2015quasiclassical}.
If the superconducting gap has nodes, the condensation energy is reduced due to low-energy quasiparticles, and the paramagnetic depairing effect is effectively strengthened because the condition $\mu_{\mathrm{B}} H\gtrsim \Delta(\bm{k})$ is easily satisfied near nodes. 
Since paramagnetic depairing suppresses the BCS state, the parity transition line $H^*(T)$ may be shifted toward lower magnetic fields and acquire a positive slope at low temperatures, as observed experimentally in $\mathrm{CeRh_2As_2}$~\cite{khanenko2025phase}.
In addition, the vortex-lattice structure can be modified.
(1-b) Material anisotropy also affects vortex properties.
However, according to previous studies~\cite{ichioka1999field}, the qualitative magnetic-field dependence of the vortex-core radius $\xi_{\mathrm{vc}}$ remains largely unchanged in the $d$-wave superconductors, although its high-field enhancement due to paramagnetic depairing in the BCS state may become more pronounced.
The LDOS reflects anisotropy most obviously. 
Fermi-surface anisotropy will lead to the distorted LDOS spectrum, while with the $d$-wave order parameter the LDOS develops long tails extending in the directions of nodes and the low-energy quasiparticle transfer is promoted between neighboring vortices~\cite{wang1995mixed,ichioka1996vortex,schopohl1995quasiparticle,ichioka1999field,ichioka1999vortex,volovik1993superconductivity}. 
Regarding $\mathrm{CeRh_2As_2}$, a fully quantitative comparison with experiment would require calculations that explicitly incorporate all anisotropies. Such calculations are beyond the scope of this paper, but the formulation provided in this paper can be extended straightforwardly for such a case by considering the momentum dependent Fermi velocity $\bm{v}_{\mathrm{F}}(\bm{k})$, normal-state density of states $N_0(\bm{k})$, and pairing function $w(\bm{k})$.
Furthermore, since both the vortex core shrinkage and the difference in the peak structures of the LDOS stem from the intrinsic interband/intraband pairing structure in the BCS and PDW states, the anisotropy of the system would retain these characteristic phenomena.
On the other hand, if the parity transition is observed in another material with an approximately isotropic electronic structure, the results of this paper would be directly applicable.
(2) Even though our quasiclassical analysis revealed the general features of vortex states in locally noncentrosymmetric superconductors, it is formally justified in the weak coupling limit. Since strong correlation effects are expected to be essential for the physical properties of $\mathrm{CeRh_2As_2}$, incorporating them into the analysis of vortex states is an important subject for future research.
(3) The existence of other orders coexisting or competing with superconductivity is also important in understanding the physical properties of $\mathrm{CeRh_2As_2}$.
The details of these orders remain elusive, and we have not yet arrived at a settled view.
This is also an interesting topic for future research. This issue can be avoided when we study multiple superconducting states at high pressures, since the coexisting order disappears there~\cite{pfeiffer2023exposing,pfeiffer2024pressure}.

\section{Acknowledgements}
We are grateful to T. Hanaguri, R. Ikeda and T. Matsushita for fruitful discussions. 
This work was supported by JSPS KAKENHI (Grant Nos. JP22H01181, JP22H04933,
JP23K17353, JP23K22452, JP24K21530, JP24H00007, JP25H01249, and
JP25KJ1479
).

%\bibliography{reference}


\appendix
\section{Derivation of effective low-energy Eilenberger equation for the bilayer Rashba model}
\label{sec:supp_1}
The Matsubara Green's function is defined as
\begin{align}
    \label{eq:sp-A1}
    \check{G}(\bm{r}_1,&\bm{r}_2,i\omega_n)\notag\\
    \equiv&-\int^{\beta}_{0}d\tau\ e^{i\omega_n\tau}\langle\hspace{-2pt}\langle T_\tau \vec{C}(\bm{r}_1,\tau)\vec{C}^{\dagger}(\bm{r}_2,0)\rangle\hspace{-2pt}\rangle\\
    =&\left(\begin{array}{cc}
        \hat{G}(\bm{r}_1,\bm{r}_2,i\omega_n)&\hat{F}(\bm{r}_1,\bm{r}_2,i\omega_n)\\
        \hat{\overline{F}}(\bm{r}_1,\bm{r}_2,i\omega_n)&\hat{\overline{G}}(\bm{r}_1,\bm{r}_2,i\omega_n)
    \end{array}\right),
\end{align}
where $\langle\hspace{-2pt}\langle\cdots\rangle\hspace{-2pt}\rangle$ denotes the grand canonical ensemble average.
$\vec{C}(\bm{r},\tau)\ (\vec{C}^{\dagger}(\bm{r},\tau))$ is the Heisenberg representation of $\vec{C}(\bm{r})\ (\vec{C}^{\dagger}(\bm{r}))$ with imaginary time $\tau$, and $T_\tau$ is the time-ordered product.
The Matsubara Green's function satisfies the Gor'kov equation
\begin{equation}
    \label{eq:sp-A2}
    \begin{aligned}
    &\left(i\omega_n\check{1}-\check{H}^N(-i\nabla_1+e\bm{A}_1)\right)\check{G}(\bm{r}_1,\bm{r}_2,i\omega_n)\\
    &-\int d^2r_3\ \check{\Delta}(\bm{r}_1,\bm{r}_3)\check{G}(\bm{r}_3,\bm{r}_2,i\omega_n)
    =\delta(\bm{r}_1,\bm{r}_2)\check{1}.
    \end{aligned}
\end{equation}
The normal Hamiltonian $\check{H}^N$ and the pair potential $\check{\Delta}$ are given in Eqs.~\eqref{eq:mm-13} and \eqref{eq:mm-14}, respectively.
In order to derive the quasiclassical equation, the gauge-covariant Wigner transform of $\check{G}$~\cite{kita2001gauge,kita2015statistical,kita2010introduction,ueki2016vortex,levanda2001wigner} is introduced by
\begin{align}
    \label{eq:sp-A5}
    \check{G}(\bm{k},\bm{r},i\omega_n)
    \equiv&\int d^2\overline{r}
    \ e^{-i\bm{k}\cdot\overline{\bm{r}}}
    \check{\Gamma}(\bm{r},\bm{r}_1)
    \check{G}(\bm{r}_1,\bm{r}_2,i\omega_n)
    \check{\Gamma}(\bm{r}_2,\bm{r})\\
    \label{eq:sp-A6}
    =&\left(\begin{array}{cc}
        \hat{G}(\bm{k},\bm{r},i\omega_n)&\hat{F}(\bm{k},\bm{r},i\omega_n)\\
        \hat{\overline{F}}(\bm{k},\bm{r},i\omega_n)&\hat{\overline{G}}(\bm{k},\bm{r},i\omega_n)
    \end{array}\right),
\end{align}
where
\begin{align}
    \label{eq:sp-A3}
    I(\bm{r}_1,\bm{r}_2)
    \equiv& e\int^{\bm{r}_2}_{\bm{r}_1}d\bm{s}\cdot\bm{A}(\bm{s}),\\
    \label{eq:sp-A4}
    \check{\Gamma}(\bm{r}_1,\bm{r}_2)
    \equiv&
    \left(\begin{array}{cc}
    e^{iI(\bm{r}_1,\bm{r}_2)}\hat{1}&0\\
    0&e^{-iI(\bm{r}_1,\bm{r}_2)}\hat{1}
    \end{array}\right).
\end{align}
The integral path in the formula of $I$ is the straight line connecting $\bm{r}_1$ to $\bm{r}_2$.
In Eq.~\eqref{eq:sp-A5}, the real-space coordinates $\bm{r}_1,\bm{r}_2$ are transformed to their center-of-mass coordinate $\bm{r}$ and relative coordinate $\overline{\bm{r}}$, so $\bm{r}_1$ and $\bm{r}_2$ on the right hand side of the equation should be regarded as $\bm{r}+\overline{\bm{r}}/2$ and $\bm{r}-\overline{\bm{r}}/2$, respectively.
With this definition, $\check{G}(\bm{k},\bm{r},i\omega_n)$ is invariant under gauge transformations with respect to $\bm{r}$.
The Wigner transform of $\check{\Delta}$ is defined in the same way.
Substituting the inverse Wigner transform of $\check{G}$ and $\check{\Delta}$ into Eq.~\eqref{eq:sp-A2} yields
\begin{equation}
    \label{eq:sp-A7}
    \check{G}^{-1}(\bm{k},\bm{r},i\omega_n)\check{G}(\bm{k},\bm{r},i\omega_n)=\check{1},
\end{equation}
where
\begin{align}
    \label{eq:sp-A8}
    &\check{G}^{-1}(\bm{k},\bm{r},i\omega_n)
    =i\omega_n\check{1}-\check{H}^{N}(\bm{k})-\check{\Delta}(\bm{k},\bm{r}),\\
    \label{eq:sp-A9}
    &\check{H}^{N}(\bm{k})
    =\left(\begin{array}{cc}
        \hat{H}^N-\frac{i}{2}\hat{\bm{v}}\cdot\bm{\partial}&0\\
        0&-\hat{H}^{N\circ}+\frac{i}{2}\hat{\bm{v}}\cdot\bm{\partial}
    \end{array}\right),\\
    \label{eq:sp-A10}
    &\check{\Delta}(\bm{k},\bm{r})
    =\left(\begin{array}{cc}
        0&\hat{\Delta}(\bm{k},\bm{r})\\
        -\hat{\Delta}^{\circ}(\bm{k},\bm{r})&0
    \end{array}\right),
\end{align}
with
\begin{align}
\label{eq:sp-A11}
    &\bm{\partial}
    =\begin{cases}
        \nabla\ \ \text{on}\ \ G,\overline{G}\\
        \nabla+2ie\bm{A}\ \ \text{on}\ \ F,\Delta\\
        \nabla-2ie\bm{A}\ \ \text{on}\ \ \overline{F},\Delta^\circ
    \end{cases},
\end{align}
and $\hat{\bm{v}}(\bm{k})=\nabla_{\bm{k}} \hat{H}^{N}(\bm{k})$.
For an arbitrary function $A(\bm{k})$, $A^\circ(\bm{k})$ stands for $A(-\bm{k})^*$. 
By taking the Hermite conjugate of Eq.~\eqref{eq:sp-A8} and performing a variable transformation $\omega_n\to -\omega_n$, the "right" Gor'kov equation
\begin{equation}
    \label{eq:sp-A12}
    \check{G}(\bm{k},\bm{r},i\omega_n)\check{G}^{-1}(\bm{k},\bm{r},i\omega_n)
    =\check{1},
\end{equation}
is obtained, where the action of the differential operator $\bm{\partial}$ on an arbitrary function $A(\bm{r})$ from the right side is defined by $A(\bm{r})\bm{\partial}=-\bm{\partial}A(\bm{r})$.

Now, we rewrite Eqs.~\eqref{eq:sp-A7} and \eqref{eq:sp-A12} in the band basis.
The eigenvalues of $\hat{K}(\bm{k})\equiv \hat{H}^N(\bm{k})-\xi(\bm{k})\tilde{\sigma}_0\otimes\tilde{\tau}_0$ are $\pm\sqrt{(t_\perp\pm h)^2+\alpha^2|\bm{g}(\bm{k})|^2}$.
That is, two Kramers-degenerate bands, each separated from the original Fermi surface by $\sqrt{t_\perp^2+\alpha^2|\bm{g}(\bm{k})|^2}$ up and down, are slightly split by the effective Zeeman energy $h_{\mathrm{eff}}(\bm{k})=h\cos\chi(\bm{k})$, where $\chi(\bm{k})$ is defined by $e^{i\chi(\bm{k})}=(t_\perp+i\alpha|\bm{g}|)/\sqrt{t_\perp^2+\alpha^2|\bm{g}|^2}$. 
We label the doubly degenerate bands with an index $\nu=1,2$ and the Zeeman splitting of each band with the pseudospin $\lambda=\pm$ as  
\begin{align}
    \label{eq:sp-A15}
    \epsilon_{\nu,\lambda}(\bm{k})
    =&(-1)^{\nu}\sqrt{(t_\perp\pm h)^2+\alpha^2|\bm{g}(\bm{k})|^2},\\
    \simeq&\epsilon_{\nu}(\bm{k})\pm (-1)^{\nu}h\cos\chi(\bm{k}),
\end{align}
where $\epsilon_{\nu}(\bm{k})=(-1)^{\nu}\sqrt{t_\perp^2+\alpha^2|\bm{g}(\bm{k})|^2}$.
Then, $\hat{K}(\bm{k})$ is diagonalized with $\hat{U}(\bm{k})\equiv (|\bm{k},1,+\rangle\ |\bm{k},1,-\rangle\ |\bm{k},2,+\rangle\ |\bm{k},2,-\rangle)$, where $|\bm{k},\nu,\lambda\rangle$ are the eigenvectors of $\hat{K}(\bm{k})$ corresponding to the eigenvalues $\epsilon_{\nu,\lambda}(\bm{k})$.
Thus, we can define the Green's function in the band basis by using $\check{U}(\bm{k})\equiv\mathrm{diag}(\hat{U},\hat{U}^{\circ})$ as
\begin{align}
    \label{eq:sp-A36}
    \check{\underline{G}}(\bm{k},\bm{r},i\omega_n)
    \equiv&\ \check{U}^{\dagger}(\bm{k})\check{G}(\bm{k},\bm{r},i\omega_n)\check{U}(\bm{k})\\
    =&
    \left(\begin{array}{cc}
        \hat{\underline{G}}(\bm{k},\bm{r},i\omega_n)
        &\hat{\underline{F}}(\bm{k},\bm{r},i\omega_n)\\
        \hat{\overline{\underline{F}}}(\bm{k},\bm{r},i\omega_n)
        &\hat{\overline{\underline{G}}}(\bm{k},\bm{r},i\omega_n)
    \end{array}\right),
\end{align}
The Gor'kov equations \eqref{eq:sp-A7}, \eqref{eq:sp-A12} are multiplied by $\check{U}^\dagger(\bm{k})$ and $\check{U}(\bm{k})$ from the left and right, respectively, to obtain
\begin{equation}
    \label{eq:sp-A24}
    \begin{aligned}
    &\check{\underline{G}}^{-1}(\bm{k},\bm{r},i\omega_n)
    \check{\underline{G}}(\bm{k},\bm{r},i\omega_n)
    =\check{1}\\
    &\quad\quad
    =\check{\underline{G}}(\bm{k},\bm{r},i\omega_n)
    \check{\underline{G}}^{-1}(\bm{k},\bm{r},i\omega_n),
    \end{aligned}
\end{equation}
where
\begin{align}
    \label{eq:sp-A25}
    &\check{\underline{G}}^{-1}(\bm{k},\bm{r},i\omega_n)
    \equiv i\omega_n\check{1}
    -\check{\underline{H}}^{N}(\bm{k})-\check{\underline{\Delta}}(\bm{k},\bm{r}),\\
    \label{eq:sp-A26}
    &\check{\underline{H}}^{N}(\bm{k})
    \equiv
    \check{\xi}(\bm{k})
    +\check{\underline{K}}(\bm{k})
    -\frac{i}{2}\check{\bm{v}}(\bm{k})\cdot\bm{\partial},\\
    &\check{\xi}(\bm{k})
    \equiv\left(\begin{array}{cc}
    \xi(\bm{k})\tilde{\sigma}_0\otimes\tilde{\tau}_0&0\\
    0&-\xi(\bm{k})\tilde{\sigma}_0\otimes\tilde{\tau}_0
    \end{array}\right),\\
    &\check{\underline{K}}(\bm{k})
    \equiv\check{U}^{\dagger}(\bm{k})\check{K}(\bm{k})\check{U}(\bm{k}),\\
    &\check{\bm{v}}(\bm{k})
    \equiv\left(\begin{array}{cc}
    \hat{\bm{v}}&0\\
    0&-\hat{\bm{v}}
    \end{array}\right),\\
    \label{eq:sp-A23}
    &\check{\underline{\Delta}}(\bm{k},\bm{r})
    \equiv\check{U}^{\dagger}(\bm{k})\check{\Delta}(\bm{k},\bm{r})\check{U}(\bm{k})\\
    &\quad\quad\quad\ 
    =\left(\begin{array}{cc}
        0&\hat{\underline{\Delta}}(\bm{k},\bm{r})\\
        -\hat{\underline{\Delta}}^{\circ}(\bm{k},\bm{r})&0
    \end{array}\right).
\end{align}

Since superconductivity is a low-energy phenomenon in the weak coupling limit ($T_{\mathrm{c}0}\ll \alpha,t_\perp\ll E_{\mathrm{F}}$), it is useful to project the equations onto the low-energy subspace in the vicinity of each band $\nu$~\cite{nagai2016multi}.
Specifically, we ignore off-diagonal matrix elements with respect to the index $\nu$, and the Green's function for the band $\nu$ is grouped together as
\begin{align}
    \label{eq:sp-A28}
    &\left(\begin{array}{cc}
        \hat{G}^{(1)}(\bm{k},\bm{r},i\omega_n)&*\\
        *&\hat{G}^{(2)}(\bm{k},\bm{r},i\omega_n)
    \end{array}\right)
    \equiv
    \check{A}
    \check{\underline{G}}(\bm{k},\bm{r},i\omega_n)
    \check{A},\\
    \label{eq:sp-A29}
    &\hat{G}^{(\nu)}(\bm{k},\bm{r},i\omega_n)
    =\left(\begin{array}{cc}
        \tilde{G}^{(\nu)}(\bm{k},\bm{r},i\omega_n)&\tilde{F}^{(\nu)}(\bm{k},\bm{r},i\omega_n)\\
        \tilde{\overline{F}}^{(\nu)}(\bm{k},\bm{r},i\omega_n)&\tilde{\overline{G}}^{(\nu)}(\bm{k},\bm{r},i\omega_n)
    \end{array}\right),
\end{align}
by using the $8\times 8$ matrix;
\begin{equation}
    \label{eq:sp-A27}
    \check{A}
    =
    \left(\begin{array}{cccc}
        \tilde{1}&0&0&0\\
        0&0&\tilde{1}&0\\
        0&\tilde{1}&0&0\\
        0&0&0&\tilde{1}
    \end{array}\right).
\end{equation}
Here, $\hat{\Delta}^{(\nu)},\,\hat{\xi}^{(\nu)},\,\hat{K}^{(\nu)},\,\hat{\bm{v}}^{(\nu)}$ are defined from $\check{\Delta},\,\check{\xi},\,\check{K},\,\check{\bm{v}}$ in the same way.
Then, the left and right Gor'kov equations for the band $\nu$ are given as
\begin{equation}
    \label{eq:sp-A30}
    \begin{aligned}
    &\hat{G}^{(\nu)-1}(\bm{k},\bm{r},i\omega_n)
    \hat{G}^{(\nu)}(\bm{k},\bm{r},i\omega_n)
    =\hat{1}\\
    &\quad\quad
    =\hat{G}^{(\nu)}(\bm{k},\bm{r},i\omega_n)
    \hat{G}^{(\nu)-1}(\bm{k},\bm{r},i\omega_n),
    \end{aligned}
\end{equation}
where
\begin{align}
    \label{eq:sp-A31}
    &\hat{G}^{(\nu)-1}(\bm{k},\bm{r},i\omega_n)
    \equiv i\omega_n\hat{1}
    -\hat{H}^{N(\nu)}(\bm{k})-\hat{\Delta}^{(\nu)}(\bm{k},\bm{r}),\\
    \label{eq:sp-A32}
    &\hat{H}^{N(\nu)}(\bm{k})
    \equiv
    \hat{\xi}^{(\nu)}(\bm{k})
    +\hat{K}^{(\nu)}(\bm{k})
    -\frac{i}{2}\hat{\bm{v}}^{(\nu)}(\bm{k})
    \cdot\bm{\partial}.
\end{align}

Subtracting the right Gor'kov equation from the left Gor'kov equation results in
\begin{align}
    &\left[i\omega_n\hat{S}_z-(-1)^{\nu}h\cos\chi\hat{T}_z-\hat{\Delta}^{(\nu)}\hat{S}_z+\frac{i}{2}\hat{\bm{v}}^{(\nu)}\cdot\bm{\partial}\hat{S}_z,\hat{S}_z\hat{G}^{(\nu)}\right]
    \notag \\ &=\hat{0},
\end{align}
where diagonal matrices $\hat{S}_z=\mathrm{diag}(1,1,-1,-1)$ and $\hat{T}_z=\mathrm{diag}(1,-1,1,-1)$ are used.
Now, we define the quasiclassical Green's function as~\cite{kopnin2001theory,kita2015statistical,ueki2016vortex}
\begin{equation}
    \label{eq:sp-A33}
    \hat{g}^{(\nu)}
    \equiv\mathcal{P}\int\frac{d\xi^{(\nu)}_{k}}{\pi}\hat{S}_zi\hat{G}^{(\nu)}
    \equiv\left(\begin{array}{cc}
        \tilde{g}^{(\nu)}&-i\tilde{f}^{(\nu)}\\
        -i\tilde{f}^{(\nu)\circ}&-\tilde{g}^{(\nu)\circ}
    \end{array}\right),
\end{equation}
where $\mathcal{P}$ denotes the principal value.
By integrating Eq.~\eqref{eq:sp-A30} over $\xi^{(\nu)}_{k}$, we reach the Eilenberger equation
\begin{equation}
    \label{eq:sp-A35}
    \begin{aligned}
        \ \Big[i\omega_n\hat{S}_z-(-1)^\nu h&\cos\chi\hat{T}_z-\hat{\Delta}^{(\nu)}\hat{S}_z,\ \hat{g}^{(\nu)}(\bm{k}_{\mathrm{F}},\bm{r},i\omega_n)\Big]\\
    &+i\bm{v}_{\mathrm{F}}\cdot\bm{\partial}\hat{g}^{(\nu)}(\bm{k}_{\mathrm{F}},\bm{r},i\omega_n)
    =\hat{0}.
    \end{aligned}
\end{equation}
Here, we used the fact that on the Fermi surface, $\hat{\bm{v}}^{(\nu)}$ is independent of $\nu$ in the first order of $\alpha/E_{\mathrm{F}}$: $\hat{\bm{v}}^{(\nu)}(\bm{k}_{\mathrm{F}})=\bm{v}(\bm{k}_{\mathrm{F}})\hat{S}_z$~\cite{dan2015quasiclassical}.
For simplicity, the momentum dependence of $\chi$ is hereinafter neglected, $\chi(\bm{k})=\chi$, where $e^{i\chi}=(t_\perp+i\alpha\langle|\bm{g}|^2\rangle^{1/2})/\sqrt{t_\perp^2+\alpha^2\langle|\bm{g}|^2\rangle}$.
The $(1,2)\ 2\times 2$ submatrix of Eq.~\eqref{eq:sp-A35} is none other than Eq.~\eqref{eq:mm-6} in the main text.
Since Eq.~\eqref{eq:sp-A35} consists only of commutator and differentiation, the normalization condition of the quasiclassical Green's function is given by $[\hat{g}^{(\nu)}]^2=\hat{1}$ as usual~\cite{nagai2016multi}.
Also, by taking a trace on both sides of Eq.~\eqref{eq:sp-A35}, the relation $\tilde{g}^{(\nu)}=\tilde{g}^{(\nu)\circ}$ is obtained, which is physically a consequence of the particle-hole symmetry~\cite{kita2015statistical,dan2015quasiclassical,kopnin2001theory}.

Up to this point, we have not assumed any specific symmetry of Cooper pairs.
Hereafter, let us consider spin-singlet intra-sublattice Cooper pairs, 
\begin{equation}
    \hat{\Delta}(\bm{k},\bm{r})
    =i\tilde{\sigma}_y\otimes
    \left(\begin{array}{cc}
        \Delta_{1}(\bm{k},\bm{r})&0\\
        0&\Delta_{2}(\bm{k},\bm{r})
    \end{array}\right).
\end{equation}
This leads to the explicit expression of $\hat{\Delta}^{(\nu)}$,
\begin{align}
    \hat{\Delta}^{(\nu)}(\bm{k},\bm{r})
    =&\left(\begin{array}{cc}
        0&\tilde{\Delta}^{(\nu)}(\bm{k},\bm{r})\\
        -\tilde{\Delta}^{(\nu)\circ}(\bm{k},\bm{r})&0
    \end{array}\right),\\
    \tilde{\Delta}^{(\nu)}(\bm{k},\bm{r})
    =&e^{-i\phi}
    \left(\begin{array}{cc}
        (-1)^{\nu+1}\sin\chi\Delta_o&-\Delta_e\\
        -\Delta_e&(-1)^{\nu+1}\sin\chi\Delta_o
    \end{array}\right),
\end{align}
where $\Delta_e\equiv(\Delta_1+\Delta_2)/2$ and $\Delta_o\equiv(\Delta_1-\Delta_2)/2$ are the sublattice-symmetric and sublattice-antisymmetic components of the order parameter.

\section{Derivation of the Adachi approximate solution}
\label{sec:supp_2}

In this section, we explain 
how Eq.~\eqref{eq:mm-1} is obtained from the Eilenberger equation in the Adachi approximation.
To simplify the notation, $\omega_n>0$ is assumed.
First, Eq.~\eqref{eq:mm-6} is transformed as
\begin{equation}
    \label{eq:sp-B1}
    \begin{aligned}
    \tilde{f}^{(\nu)}
    =&
    \{(2\omega_n+i\bm{v}_{\mathrm{F}}\cdot\bm{\Pi})\tilde{\upsilon}_0
    +i(-1)^\nu h\cos\chi[\tilde{\upsilon}_z,\cdot]
    \}^{-1}\\
    &\times\{\tilde{\Delta}^{(\nu)}
    \tilde{g}^{(\nu)}
    +\tilde{g}^{(\nu)}
    \tilde{\Delta}^{(\nu)}\}.
    \end{aligned}
\end{equation}
Expansion of $\Delta_j$ in terms of the Landau levels [Eq.~\eqref{eq:mm-10}] leads to
\begin{equation}
    \tilde{\Delta}^{(\nu)}(\bm{k}_{\mathrm{F}},\bm{r})
    =e^{-i\phi(\bm{k}_{\mathrm{F}})}w(\bm{k}_{\mathrm{F}})
    \sum_{N}\Psi_N(\bm{r})\tilde{d}^{(\nu)}_{N},
\end{equation}
where
\begin{equation}
    \tilde{d}^{(\nu)}_{N}
    =
    \left(\begin{array}{cc}
        (-1)^{\nu+1}\sin\chi d_{o,N}&-d_{e,N}\\
        -d_{e,N}&(-1)^{\nu+1}\sin\chi d_{o,N}
    \end{array}\right),
\end{equation}
is the coefficient matrix of $\Psi_N$.

As is often done, an auxiliary variable $\rho$ is introduced to express the inverse operator in the form of the integral of an exponential function;
\begin{equation}
\begin{aligned}
    \{(2&\omega_n+i\bm{v}_{\mathrm{F}}\cdot\bm{\Pi})\tilde{\upsilon}_0
    +i(-1)^\nu h\cos\chi[\tilde{\upsilon}_z,\cdot]
    \}^{-1}\\
    =&
    \int^{\infty}_{0}d\rho\ 
    e^{-2\omega_n\rho}
    e^{-i(-1)^{\nu}\rho h\cos\chi[\tilde{\upsilon}_z,\cdot]}
    e^{-i\rho\bm{v}_{\mathrm{F}}\cdot\bm{\Pi}}.
\end{aligned}
\end{equation}
Now, we define the ladder operators of the Landau levels corresponding to the gauge $\bm{A}(\bm{r})=Hx\hat{\bm{y}}$ as
\begin{equation}
    a\equiv
    \frac{r_H}{\sqrt{2}}
    (\Pi_y+i\Pi_x),
    \quad
    a^{\dagger}\equiv
    \frac{r_H}{\sqrt{2}}
    (\Pi_y-i\Pi_x).
\end{equation}
Since the commutation relation $[a,a^\dagger]=1$ holds, the Baker-Campbell-Hausdorff formula derives
\begin{equation}
    e^{-i\rho\bm{v}_{\mathrm{F}}\cdot\bm{\Pi}}
    =e^{-i(s^*\rho a^\dagger+s\rho a)}
    =e^{-|s|^2\rho^2/2}
    e^{-is^*\rho a^\dagger}
    e^{-is\rho a},
\end{equation}
where $s=v_{\mathrm{F}}(\hat{k}_y-i\hat{k}_x)/(\sqrt{2}r_H)$.
Using that $a,a^\dagger$ lowers and raises the Landau level as $a\Psi_N=\sqrt{N}\Psi_{N-1},a^\dagger\Psi_N=\sqrt{N+1}\Psi_{N+1}$, the action of $e^{-i\rho\bm{v}_{\mathrm{F}}\cdot\bm{\Pi}}$ on $\Psi_N$ is given by
\begin{equation}
    \label{eq:sp-B2}
    e^{-i\rho\bm{v}_{\mathrm{F}}\cdot\bm{\Pi}}
    \Psi_N(\bm{r})
    =e^{-|s|^2\rho^2/2}
    \sum_{M}\Psi_M(\bm{r})
    \mathcal{L}_{MN}(-is^*\rho),
\end{equation}
where
\begin{equation}
    \mathcal{L}_{MN}(z)
    =
    \sum_{l=0}^{\mathrm{min}(M,N)}
    \frac{\sqrt{M!N!}}{(M-l)!(N-l)!l!}
    (z)^{M-l}(-z^*)^{N-l}.
\end{equation}
By substituting Eq.~\eqref{eq:sp-B2} into the definition of $\tilde{\Phi}^{(\nu)}$ [Eq.~\eqref{eq:mm-7}], Eq.~\eqref{eq:mm-1} follows.

For the general form of the order parameters, solving Eq.~\eqref{eq:mm-8} in step (ii) of the self-consistent calculation must rely on a numerical calculation.
But, if either $\Delta_e$ or $\Delta_o$ is absent, it can be carried out analytically: the solutions are given as, for the BCS state,
\begin{equation}
\label{eq:res-1}
\begin{cases}
    g_{++}^{(\nu)}=&\pm \frac{1+a_1-a_2}{\sqrt{(1-a_1-a_2)^2-4a_1a_2}},\\
    g_{--}^{(\nu)}=&\pm \frac{1-a_1+a_2}{\sqrt{(1-a_1-a_2)^2-4a_1a_2}},\\
    g_{+-}^{(\nu)}=&g_{-+}^{(\nu)}=0,
    \end{cases}
\end{equation}
where $a_1=\Phi^{(\nu)}_{+-}\Phi^{(\nu)\circ}_{-+},\ a_2=\Phi^{(\nu)}_{-+}\Phi^{(\nu)\circ}_{+-}$, and for the PDW state, 
\begin{equation}
\label{eq:res-2}
\begin{cases}
    g_{++}^{(\nu)}=&\pm (1-4b_1)^{-1/2},\\
    g_{--}^{(\nu)}=&\pm (1-4b_2)^{-1/2},\\
    g_{+-}^{(\nu)}=&g_{-+}^{(\nu)}=0,
    \end{cases}
\end{equation}
where $b_1=\Phi^{(\nu)}_{++}\Phi^{(\nu)\circ}_{++},\ b_2=\Phi^{(\nu)}_{--}\Phi^{(\nu)\circ}_{--}$.
The signs of $g^{(\nu)}_{++},\,g^{(\nu)}_{--}$ must be chosen to be positive, due to the requirement of a smooth connection to the spatially homogeneous state at infinity.

\section{Derivation of gap equation}
\label{sec:supp_3}

We assume the separable pairing interaction, $V(\bm{k},\bm{k}')=-Vw(\bm{k})w^*(\bm{k}')\ (V>0)$.
In the mean-field approximation, the order parameter is defined in terms of electron operators by
\begin{equation}
    \label{eq:sp-C1}
    \Delta_{l}(\bm{q},\bm{k})
    =-\frac{V}{2}w(\bm{k})\langle\hspace{-2pt}\langle \Psi_{\bm{q},l}\rangle\hspace{-2pt}\rangle,
\end{equation}
where
\begin{equation}
  \Psi_{\bm{q},l}
  =\frac{1}{\Omega}
  \sum_{\bm{k},s_1,s_2}
  w^*(\bm{k})c_{-\bm{k}+\bm{q}/2,s_1,l}(-i\tilde{\sigma}_y)_{s_1s_2}c_{\bm{k}+\bm{q}/2,s_2,l},
\end{equation}
is the field operator of the Cooper pair with the total momentum $\bm{q}$ in the layer $l$, and $\Omega$ is the area of each layer.
The Wigner transform of $\Delta_e,\Delta_o$ is written as
\begin{align}
    \label{eq:sp-C2}
    \Delta_e(\bm{k},\bm{r})
    =&\frac{VT}{4\Omega}
    w
    \sum_{\bm{k}',|\omega_n|<\varepsilon_{\rm c}}
    w^*\mathrm{Tr}\left[
    (i\tilde{\sigma}_y\otimes \tilde{\tau}_0)
    \hat{F}(\bm{k}',\bm{r},i\omega_n)\right],\\
    \label{eq:sp-C3}
    \Delta_o(\bm{k},\bm{r})
    =&\frac{VT}{4\Omega}
    w
    \sum_{\bm{k}',|\omega_n|<\varepsilon_{\rm c}}
    w^*\mathrm{Tr}\left[
    (i\tilde{\sigma}_y\otimes \tilde{\tau}_z)
    \hat{F}(\bm{k}',\bm{r},i\omega_n)\right],
\end{align}
with the anomalous Green's function $\hat{F}$, where $\varepsilon_{\rm c}$ is the cut-off energy to prevent logarithmic divergence.
Here, let $V$ be $j$-dependent, since amplitudes of the attractive interaction can be different for $\Delta_e$ and $\Delta_o$ when considering, e.g., multipole-fluctuation-induced interactions~\cite{nogaki2024field}.
We rewrite Eqs.~\eqref{eq:sp-C2} and \eqref{eq:sp-C3} in the band basis as done in Appendix~\ref{sec:supp_1}, and performing the integration over the energy yields
\begin{align}
    \label{eq:sp-C4}
    \frac{\Delta_e}{V_e}
    =&-\frac{\pi T}{4}w
    \sum_{\nu,|\omega_n|<\varepsilon_{\rm c}}N_{0}^{(\nu)}
    \left\langle
        w^*
        e^{i\phi}
        \mathrm{Tr}\left[\tilde{\upsilon}_x
        \tilde{f}^{(\nu)}\right]
    \right\rangle,\\
    \label{eq:sp-C5}
    \frac{\Delta_o}{V_o}
    =&-\frac{\pi T}{4}w
    \sum_{\nu,|\omega_n|<\varepsilon_{\rm c}}N_{0}^{(\nu)}
    (-1)^{\nu}\sin\chi
    \left\langle
        w^*
        e^{i\phi}
        \mathrm{Tr}\left[
        \tilde{f}^{(\nu)}\right]
    \right\rangle.
\end{align}
Since $\delta N\equiv |N_{0}^{(1)}-N_{0}^{(2)}|/(N_{0}^{(1)}+N_{0}^{(2)})$ is of the order $\sqrt{\alpha^2+t_\perp^2}/E_{\mathrm{F}}$, the limit $\delta N\to 0\ (N_{0}^{(1)}=N_{0}^{(2)}=N_0)$ is assumed in the following.
The cut-off energy $\varepsilon_{\rm c}$ can be eliminated by using the well-known relations,
\begin{align}
    \label{eq:sp-C6}
    \frac{1}{V_e}
    =&
    \sum_{\nu}\frac{N_0}{2}\left(\ln\frac{T}{T_{\mathrm{c}0}^{e}}+\pi T\sum_{|\omega_n|<\varepsilon_{\rm c}}\frac{1}{|\omega_n|}\right),\\
    \label{eq:sp-C7}
    \frac{1}{V_o}
    =&
    \sin^2\chi
    \sum_{\nu}\frac{N_0}{2}
    \left(\ln\frac{T}{T_{\mathrm{c}0}^{o}}
    +\pi T\sum_{|\omega_n|<\varepsilon_{\rm c}}
    \frac{1}{|\omega_n|}\right),
\end{align}
where $T_{\mathrm{c}0}^{e}$ and $T_{\mathrm{c}0}^{o}$ are the transition temperature of the BCS and PDW state at zero magnetic field, respectively.
Finally, substituting the expansion [Eq.~\eqref{eq:mm-10}] into Eqs.~\eqref{eq:sp-C4}, \eqref{eq:sp-C5} and using the orthogonal normality of $\{\Psi_N\}$, i.e., $\overline{\Psi_M^*\Psi_N}=\delta_{M,N}$, we get Eqs.~\eqref{eq:mm-2}, \eqref{eq:mm-3}.

\section{Derivation of condensation energy}
\label{sec:supp_4}

Generally, based on the coupling constant integration method~\cite{graser2004influence,abrikosov2012methods,uematsu2019chiral,vorontsov2003thermodynamic,vorontsov2005phase,serene1983quasiclassical,sauls2024fermi}, the condensation energy of the superconducting state is calculated by
\begin{equation}
\label{eq:sp-D1}
  \delta\mathcal{F}\equiv\mathcal{F}_S-\mathcal{F}_N=\int^{1}_{0}\frac{dg}{g}
  \langle\hspace{-2pt}\langle g\mathcal{H}_{\mathrm{int}}\rangle\hspace{-2pt}\rangle_{g},
\end{equation}
where $\mathcal{H}_{\mathrm{int}}$ is the interacting part of the Hamiltonian and $\langle\hspace{-2pt}\langle\cdots\rangle\hspace{-2pt}\rangle_g=\mathrm{tr}[e^{-\beta \mathcal{H}(g)}\cdots]/\mathrm{tr}[e^{-\beta \mathcal{H}(g)}]$ is the grand canonical ensemble average with respect to $\mathcal{H}(g)=\mathcal{H}_0+g\mathcal{H}_{\mathrm{int}}$. Here, $\mathcal{H}_0$ is the non-interacting Hamiltonian.
For the bilayer Rashba model, $\delta\mathcal{F}$ is expressed with the quasiclassical Green's function in the band basis as
\begin{equation}
    \label{eq:sp-D2}
    \begin{aligned}
    \delta\mathcal{F}
    =&
    \frac{\pi TN_0}{2}
    \int^{1}_{0}\frac{dg}{g}\int d^2r
    \sum_{\nu,|\omega_n|<\varepsilon_{\rm c}}
    \mathrm{Tr}
    \langle \tilde{f}^{(\nu)\circ}(g)
    \tilde{\Delta}^{(\nu)}(g)\\
    &+\tilde{\Delta}^{(\nu)\circ}(g)
    \tilde{f}^{(\nu)}(g)
    \rangle,
    \end{aligned}
\end{equation}
where $\tilde{f}^{(\nu)}(\bm{k}_{\mathrm{F}},\bm{r},i\omega_n;g)$ and $\tilde{\Delta}^{(\nu)}(\bm{k}_{\mathrm{F}},\bm{r};g)$ are the self-consistent solution of Eqs.~\eqref{eq:sp-A35}, \eqref{eq:sp-C4}, \eqref{eq:sp-C5} with the amplitude of the interaction being $gV_j\ (0\leq g\leq 1)$.
Now, integration over the amplitude of the interaction is substituted by integration over the pair potential.
We introduce $x_j(g)$ by the relation
\begin{equation}
    \label{eq:sp-D3}
    x_j(g)\Delta_j(\bm{k}_{\mathrm{F}},\bm{r})
    =\Delta_j(\bm{k}_{\mathrm{F}},\bm{r};g),
\end{equation}
for $j=e,o$, and assume that the inverse function $g=g_j(x_j)$ is obtained.
$\tilde{f}^{(\nu)}(x_j)$ is now the (non-self-consistent) solution of Eq.~\eqref{eq:sp-A35} with replacing $\Delta_j$ with $x_j\Delta_{j,\mathrm{sc}}$, where $\Delta_{j,\mathrm{sc}}$ is the self-consistent solution of Eqs.~\eqref{eq:sp-A35}, \eqref{eq:sp-C4}, \eqref{eq:sp-C5}.
By substituting Eq.~\eqref{eq:sp-D3} into the gap equation [Eqs.~\eqref{eq:sp-C4} and \eqref{eq:sp-C5}] and differentiating them with respect to $x_j$, we find the formulae
\begin{align}
    \label{eq:sp-D4}
    &\begin{aligned}
    \frac{dg_e}{dx_e}\frac{\Delta_e(x_e)}{g_e}
    =&\Delta_e
    +\frac{\pi TN_0}{4}\sum_{\nu,|\omega_n|<\varepsilon_{\rm c}}\\
    \times&
    g_eV_ew\langle w^*e^{i\phi}
    \mathrm{Tr}[\tilde{\upsilon}_x\partial_{x_e}\tilde{f}^{(\nu)}(x_e)]\rangle,
    \end{aligned}\\
    \label{eq:sp-D5}
    &\begin{aligned}
    \frac{dg_o}{dx_o}\frac{\Delta_o(x_o)}{g_o}
    =&\Delta_o
    +\frac{\pi TN_0}{4}\sum_{\nu,|\omega_n|<\varepsilon_{\rm c}}
    (-1)^{\nu}\sin\chi\\
    \times&
    g_oV_ow\langle w^*e^{i\phi}
    \mathrm{Tr}[\partial_{x_o}\tilde{f}^{(\nu)}(x_o)]\rangle.
    \end{aligned}
\end{align}
Inserting Eqs.~\eqref{eq:sp-D4} and \eqref{eq:sp-D5} to Eq.~\eqref{eq:sp-D2} leads to 
\begin{widetext}
\begin{align}
    \label{eq:sp-D6-1}
    \delta\mathcal{F}
    =&\delta\mathcal{F}_e+\delta\mathcal{F}_o,\\
    \label{eq:sp-D6-2}
    \delta\mathcal{F}_e
    =&
    \pi TN_0
    \int d^2r
    \sum_{\nu,|\omega_n|<\varepsilon_{\rm c}}
    \mathrm{Re}\left(
    \langle \Delta_{e}^{\circ}e^{i\phi(-\bm{k}_{\mathrm{F}})}
    \mathrm{Tr}[\tilde{\upsilon}_x\tilde{f}^{(\nu)}]\rangle
    -2\int^{1}_{0}dx_e
    \langle \Delta_{e}^{\circ}e^{i\phi(-\bm{k}_{\mathrm{F}})}
    \mathrm{Tr}[\tilde{\upsilon}_x\tilde{f}^{(\nu)}(x_e)]\rangle
    \right),\\
    \label{eq:sp-D6-3}
    \delta\mathcal{F}_o
    =&
    \pi TN_0
    \int d^2r
    \sum_{\nu,|\omega_n|<\varepsilon_{\rm c}}
    (-1)^{\nu}\sin\chi
    \mathrm{Re}\left(
    \langle \Delta_{o}^{\circ}e^{i\phi(-\bm{k}_{\mathrm{F}})}
    \mathrm{Tr}[\tilde{f}^{(\nu)}]\rangle
    -2\int^{1}_{0}dx_o
    \langle \Delta_{o}^{\circ}e^{i\phi(-\bm{k}_{\mathrm{F}})}
    \mathrm{Tr}[\tilde{f}^{(\nu)}(x_o)]\rangle\right),
\end{align}
\end{widetext}
where the gap equations \eqref{eq:sp-C4} and \eqref{eq:sp-C5} are used and a partial integration with respect to $x_j$ is performed on the second term of $\delta\mathcal{F}_j$.
The cut-off energy $\varepsilon_{\rm c}$ is eliminated by employing Eqs.~\eqref{eq:sp-C4}, \eqref{eq:sp-C5}, \eqref{eq:sp-C6}, \eqref{eq:sp-C7}, and at long last, the expression of Eq.~\eqref{eq:mm-4} is reached.

\subsection{Expansion with respect to the order parameter}

We can expand the condensation energy $\delta\mathcal{F}$ with respect to the order parameter when regions near the upper critical field are considered.
The solution of the Eilenberger equation with perturbative expansion is given as $\tilde{f}^{(\nu)}\simeq 2\,\mathrm{sgn}(\omega_n)\tilde{\Phi}_\nu$ up to the first order of $\Delta$, where $\mathrm{sgn}(\omega_n)=\omega_n/|\omega_n|$.
For the $s$-wave case ($w(\bm{k})=1$), substituting this solution into Eq.~\eqref{eq:sp-D6-1} yields the quadratic term of $\delta\mathcal{F}$ as
\begin{widetext}
\begin{align}
    \delta\mathcal{F}^{(2)}
    =&4\Omega\sum_{\nu}\frac{N_0}{2}
    \sum_{j,N}
    |d_{j,N}|^2E_{j,N},\\
    E_{e,N}
    =&\ln\frac{T}{T_{\mathrm{c}0}^{e}}
    +2\pi T
    \int^{\infty}_{0}\frac{d\rho}{\sinh(2\pi T\rho)}\left(1-e^{-|s|^2\rho^2/2}\cos(2\rho h\cos\chi)L_N\left(\frac{v_{\mathrm{F}}^2\rho^2}{2r_H^2}\right)\right),\\
    E_{o.N}
    =&\sin^2\chi
    \left\{
    \ln\frac{T}{T_{\mathrm{c}0}^{o}}
    +2\pi T
    \int^{\infty}_{0}\frac{d\rho}{\sinh(2\pi T\rho)}\left(1-e^{-|s|^2\rho^2/2}L_N\left(\frac{v_{\mathrm{F}}^2\rho^2}{2r_H^2}\right)\right)\right\},
\end{align}
\end{widetext}
with using the formula
\begin{equation}
  \langle \mathcal{L}_{MN}(-is^*\rho)\rangle
  =L_N\left(\rho^2\frac{v_{\mathrm{F}}^2}{2r_H^2}\right)\delta_{MN},
\end{equation}
where 
\begin{equation}
  L_N(x)=\sum_{l=0}^{N}\frac{N!}{(l!)^2(N-l)!}(-x)^{l},
\end{equation}
is the $N$th Laguerre polynomial.
Since $E_{j,N}$ is negative (positive) in the superconducting (normal) state, the condition $E_{j,N}=0$ corresponds to the criteria of the normal-superconducting phase transition. 

\section{Derivation of local density of states}
\label{sec:supp_5}

\begin{figure*}[tbp]
  \centering
  \includegraphics[width=0.75\textwidth]{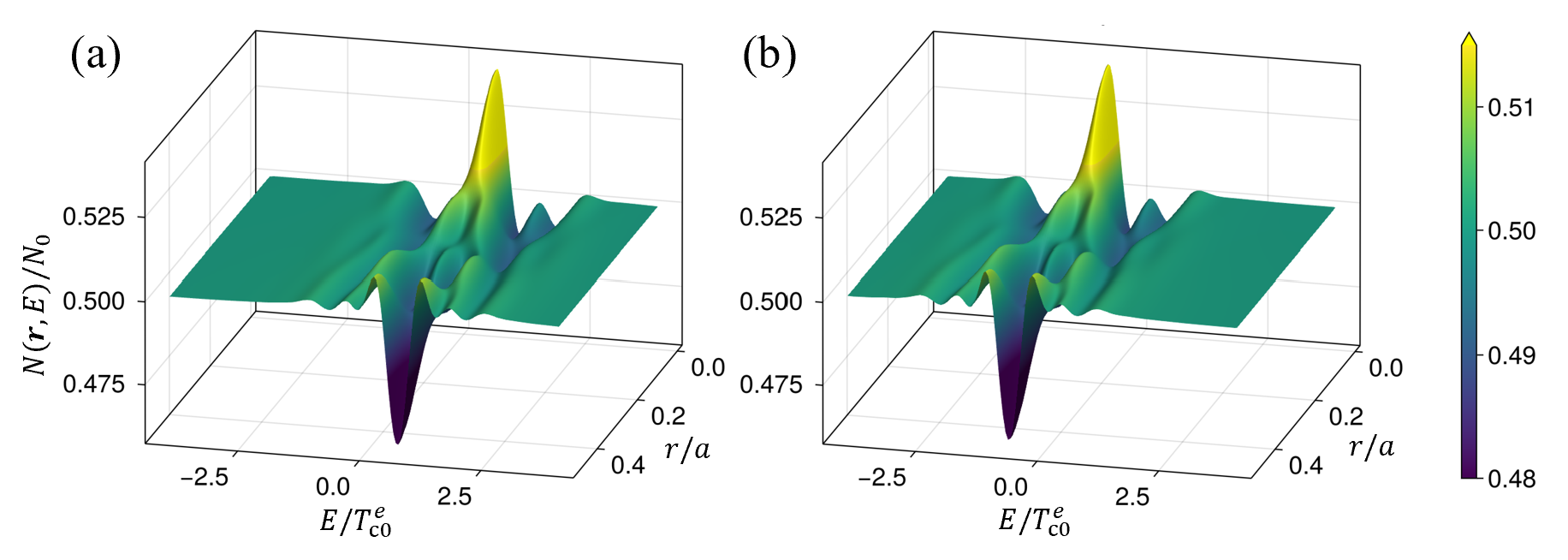}
  \caption{The spin-resolved LDOS near the vortex core in the BCS state.
  Panels (a) and (b) show the up and down spin components of the LDOS, respectively.
  The temperature and magnetic field are fixed at $T/T_{\mathrm{c}0}^{e}=0.74,\ \mu_{\mathrm{B}} H/T_{\mathrm{c}0}^{e}=10.25$, as in Figs.~\ref{fig:ldos_bcs_pdw} and \ref{fig:ldos_meron}.
  }
\label{fig:spin_resolved_ldos}
\end{figure*}

The retarded Green's function is defined as
\begin{align}
    \check{G}^{R}(\bm{r}_1,&\bm{r}_2,t_1-t_2)\notag\\
    \equiv&-i\theta(t_1-t_2)\langle\hspace{-2pt}\langle \{\vec{C}(\bm{r}_1,t_1),\vec{C}^{\dagger}(\bm{r}_2,t_2)\}\rangle\hspace{-2pt}\rangle,\\
    \check{G}^{R}(\bm{r}_1,&\bm{r}_2,E)
    \equiv\int^{\infty}_{-\infty}dt\ e^{iEt}
    \check{G}^{R}(\bm{r}_1,\bm{r}_2,t),
\end{align}
where $\vec{C}(\bm{r},t)\ (\vec{C}^{\dagger}(\bm{r},t))$ denotes the real-time Heisenberg representation of $\vec{C}(\bm{r})\ (\vec{C}^{\dagger}(\bm{r}))$, $\theta(t)$ is the Heaviside step function, and $\{A,B\}=AB+BA$.
The retarded Green's function is related to the Matsubara Green's function via an analytic continuation,
\begin{equation}
    \check{G}^{R}(E)
    =
    \check{G}(i\omega_n\to E+i\eta),
\end{equation}
where the parameter $\eta>0$ represents a positive infinitesimal.

By performing the Wigner transformation in the same manner as in Eq.~\eqref{eq:sp-A5}, we obtain $\check{G}^{R}(\bm{k},\bm{r},E)$.
Then, the LDOS $N(\bm{r},E)$ is calculated from it as 
\begin{equation}
    N(\bm{r},E)
    =
    -\frac{1}{\pi\Omega}
    \sum_{\bm{k}}\mathrm{Im}
    \mathrm{Tr}[\hat{G}^{R}(\bm{k},\bm{r},E)],
\end{equation}
where $\hat{G}^{R}$ is the normal retarded Green's function, i.e., the (1,1) 4$\times$4 submatrix of $\check{G}^{R}$.
Applying sequentially the same unitary transformations as in Eqs.~\eqref{eq:sp-A36} and \eqref{eq:sp-A28} yields
\begin{equation}
    \label{eq:sp-F1}
    N(\bm{r},E)
    =
    -\frac{1}{\pi\Omega}
    \sum_{\bm{k},\nu}\mathrm{Im}
    \mathrm{Tr}[\tilde{G}^{(\nu)R}(\bm{k},\bm{r},E)].
\end{equation}
Furthermore, the summation over the momentum space in the right hand side is decomposed into an energy integral and an angular average over the Fermi surface as
\begin{equation}
    \sum_{\bm{k}}\cdots
    \simeq\frac{N_0}{2}\Omega\int d\xi_{k}^{(\nu)}\langle\cdots\rangle.
\end{equation}
Here, we assume that the normal-state density of states per unit area at the Fermi level, $N_{0}^{(\nu)}(\bm{k}_{\mathrm{F}})$, is isotropic and identical for the two bands $\nu=1,2$.
The denominator $2$ arises from the fact that each band is doubly degenerate.
The energy integral of the Green's function in Eq.~\eqref{eq:sp-F1} leads to the quasiclassical Green's function, and results in Eq.~\eqref{eq:mm-15}.

To evaluate the spin- and sublattice-resolved LDOS, the retarded quasiclassical Green's function needs to be transformed from the band basis to the spin-sublattice basis.
Specifically, we apply the inverse transformation of Eq.~\eqref{eq:sp-A36} to $\tilde{g}^{(\nu)R}(\bm{k},\bm{r},E)$ as
\begin{equation}
    \hat{g}^{R}(\bm{k},\bm{r},E)
    =\hat{U}(\bm{k})
    \left(\begin{array}{cc}
    \tilde{g}^{(1)R}&0\\
    0&\tilde{g}^{(2)R}
    \end{array}\right)
    \hat{U}^{\dagger}(\bm{k}).
\end{equation}
Each spin-sublattice component of the LDOS is then obtained as
\begin{equation}
    N_{sl}(\bm{r},E)
    =\frac{N_0}{2}
    \mathrm{Re}\langle g^{R}_{sl}(\bm{k}_{\mathrm{F}},\bm{r},E)\rangle,
\end{equation}
where $g^{R}_{sl}$ denotes the diagonal element of $\hat{g}^{R}$ for spin $s$ and sublattice $l$.
Finally, the spin-resolved and sublattice-resolved LDOS are given by
\begin{equation}
    N_{s}(\bm{r},E)=\sum_{l}N_{sl}(\bm{r},E),
    \quad
    N_{l}(\bm{r},E)=\sum_{s}N_{sl}(\bm{r},E),
\end{equation}
respectively.
Since unitary transformations preserve the trace, the relation
\begin{equation}
    N(\bm{r},E)=\sum_{s,l}N_{sl}(\bm{r},E),
\end{equation}
follows immediately.

Figure~\ref{fig:spin_resolved_ldos} shows the spin-resolved LDOS of the BCS state for (a) up-spin and (b) down-spin quasiparticles, respectively.
The temperature and magnetic field are fixed at $T/T_{\mathrm{c}0}^{e}=0.74,\ \mu_{\mathrm{B}} H/T_{\mathrm{c}0}^{e}=10.25$, as in Figs.~\ref{fig:ldos_bcs_pdw} and \ref{fig:ldos_meron}.
The spectral peak shifts from zero energy toward the positive (negative) energy side for the up-spin (down-spin) quasiparticles.
Its position, $E/T_{\mathrm{c}0}^{e}\simeq\pm 0.70$, corresponds to the effective Zeeman field $h_{\mathrm{eff}}=\mu_{\mathrm{B}}H \cos\chi=0.68\,T_{\mathrm{c}0}^{e}$.
In each panel, the LDOS spectrum is symmetric about the respective shifted peak position.

\section{Reliability of self-consistent calculations for Eilenberger equation}
\label{sec:supp_6}

In this appendix, we examine the reliability of self-consistent calculations used to solve the Eilenberger equation [Eq.~\eqref{eq:mm-6}], and subsequently discuss the possible origins of discrepancy between our theoretical results and experimental observations.
The following three major assumptions are made in our analytical and numerical calculations: (1) The Landau level expansion is truncated at a finite order $N_{\mathrm{max}}$, (2) The Matsubara frequencies are truncated at a finite value $\varepsilon_{\mathrm{c}}$, and (3) The Adachi approximation is employed to solve the Eilenberger equation.
In the following, we assess the validity of these.

\begin{figure}[tbp]
  \centering
  \includegraphics[width=0.48\textwidth]{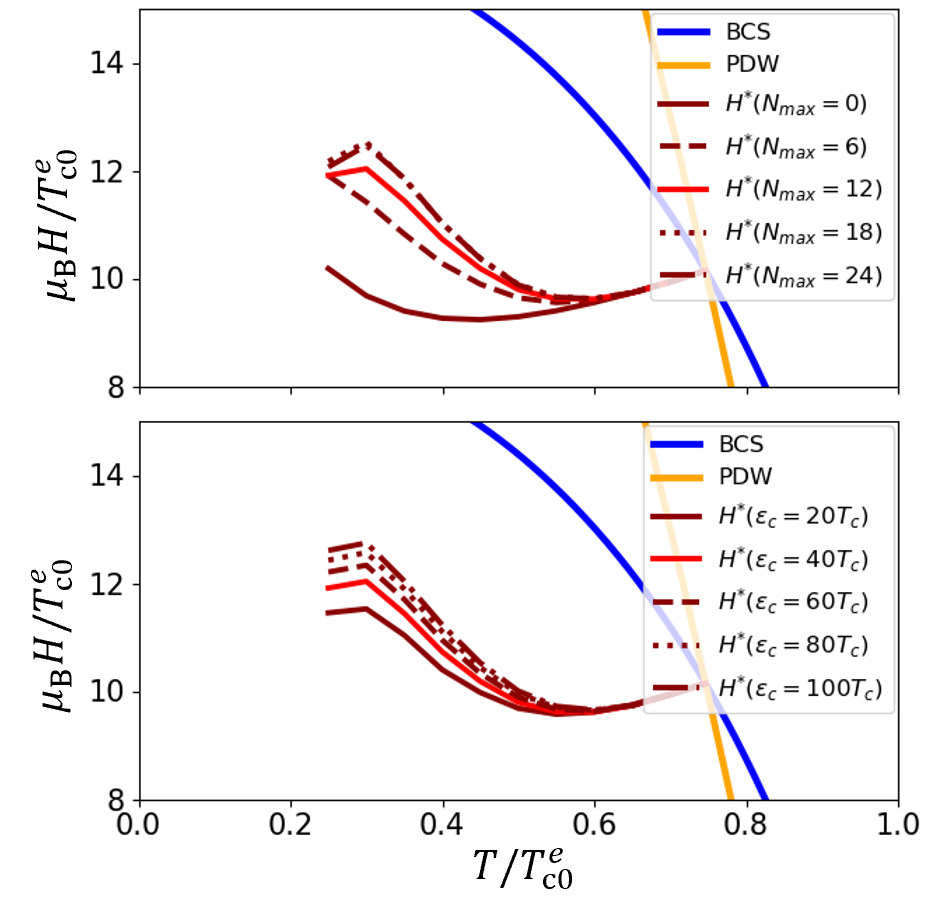}
  \caption{
  The parity transition lines from self-consistent calculations of the Eilenberger equation under various Landau level cut-offs $N_{\mathrm{max}}$ and the Matsubara frequency cut-offs $\varepsilon_{\mathrm{c}}$.
  (a) The red solid, dark red solid, dashed, dotted and dash-dotted lines represent the parity transition lines calculated with the Landau level cut-off $N_{\mathrm{max}}=12,0,6,18,24$, respectively, while $\varepsilon_{\mathrm{c}}/T_{\mathrm{c}0}^{e}$ is fixed at $40$.
  (b) The red solid, dark red solid, dashed, dotted and dash-dotted lines represent the parity transition lines calculated with the Matsubara frequency cut-off $\varepsilon_{\mathrm{c}}/T_{\mathrm{c}0}^{e}=40,20,60,80,100$, respectively, while $N_{\mathrm{max}}$ is fixed at $12$.
  }
\label{fig:self_consistent_convergence}
\end{figure}

First, we investigate the effects of truncating the Landau levels and Matsubara frequencies by varying their cut-offs and comparing the resulting numerical solutions.
In general, increasing these cut-offs is expected to improve quantitative accuracy.
In particular, while the upper critical field is typically described solely by the lowest Landau level, higher Landau levels become increasingly important away from $H_{\mathrm{c}2}$.
Figure~\ref{fig:self_consistent_convergence} shows the parity transition line $H^*(T)$ obtained from self-consistent calculations using different cut-offs.
Note that the scale of the vertical axis is enlarged compared to Figure~\ref{fig:phase_diagram}.
In Fig.~\ref{fig:self_consistent_convergence}(a), $\varepsilon_{\mathrm{c}}$ is fixed at $40\,T_{\mathrm{c}0}^{e}$ and $N_{\mathrm{max}}$ is varied from $0$ to $24$.
In Fig.~\ref{fig:self_consistent_convergence}(b), $N_{\mathrm{max}}=12$ is fixed while $\varepsilon_{\mathrm{c}}$ is changed from $20\,T_{\mathrm{c}0}^{e}$ to $100\,T_{\mathrm{c}0}^{e}$.
In both cases, incorporating more Landau levels or Matsubara frequencies shifts the transition line slightly toward higher magnetic fields.
Importantly, however, the qualitative behavior of $H^*(T)$ remains unchanged, and the magnitude of the shift gradually diminishes.
Based on this observation, we concluded that the self-consistent calculation converges with respect to both the Landau level and the Matsubara frequency.
We also concluded that the parameters $N_{\mathrm{max}}=12,\ \varepsilon_{\mathrm{c}}=40\,T_{\mathrm{c}0}^{e}$, used throughout the main text, produce results sufficiently close to the converged value.

Next, we discuss the applicability of Adachi's "approximate solution".
As mentioned in Sec.~\ref{sec:model_and_method}, this approximation, first introduced by Adachi \textit{et al.}~\cite{adachi2005anisotropic,adachi2006basal,dan2015quasiclassical, adachi2005mixed}, is an extension of the Pesch approximation~\cite{pesch1975density}.
The Pesch approximation (or the Brandt-Pesch-Tewordt approximation~\cite{brandt1967theory}) is well suited for treating vortex lattice states.
Its central assumption is that the normal Green's function is nearly uniform in high magnetic fields.
The anomalous Green's function depends linearly on the order parameter $\Delta$, whereas the leading contribution of $\Delta$ to the normal Green's function is quadratic, as can be seen from the normalization condition $g^2-ff^{\circ}=1$.
Consequently, when the order parameter is small [$\Delta\ll \omega_n$], the spatial variations of the normal Green's function are strongly suppressed.
Based on this observation, Pesch obtained an analytical solution of the Eilenberger equation by replacing the normal Green's function and the magnetic flux density with their spatial averages and by restricting the order parameter to the lowest Landau level.
Adachi \textit{et al.} extended this approach by partially taking the spatial variation of the normal Green's function into account.
Specifically, the terms proportional to $g(\bm{r})$ in the formula are retained, while the terms proportional to $\bm{\partial}g(\bm{r})$ are ignored when compared to $\bm{\partial}\Delta(\bm{r})$.
The formula was also extended to take into account contributions from higher Landau levels.
Although both approximations were originally expected to be valid only in the vicinity of the upper critical field $H\lesssim H_{\mathrm{c}2}$, subsequent numerical studies have demonstrated that they show minor derivations from the exact solutions over a wide temperature-magnetic field range~\cite{kusunose2004quasiclassical,adachi2006basal}.
This applicability is further supported by the fact that both approximations reproduce the spatially uniform BCS theory in the zero-field limit.

Finally, we address the problem that self-consistent solutions cannot be obtained at low temperatures in the present calculations.
In this study, we have performed self-consistent iterations at each temperature to determine the parity transition line $H^*(T)$. However, convergence becomes worse as the temperature decreases, and the convergence criterion is no longer satisfied below $T/T_{\mathrm{c}0}^{e}=0.25$.
One possible origin is an insufficient cut-off in the Landau levels or Matsubara frequencies.
However, as confirmed above, the influence of the cut-off is limited down to moderate temperatures (Fig.~\ref{fig:self_consistent_convergence}), and actually increasing the cut-off values does not improve the convergence at low temperatures.
This suggests that the cut-off in the numerical calculations is not the primary cause of the convergence failure.
A more plausible explanation is the breakdown of the "approximate solution" away from the upper critical field.
In deed, we find that self-consistent calculations converge even below $T/T_{\mathrm{c}0}^{e}=0.25$ when the magnetic field is close to $H_{\mathrm{c}2}$.
Therefore, the poor convergence is not a low-temperature effect, but rather occurs in regions far from the upper critical field, where the validity of the approximation deteriorates.
We thus attribute the failure of convergence primarily to the limitations of the "approximate solution".
Based on this analysis, we conclude that the discrepancy between the slope of the parity transition line $H^*(T)$ obtained in our quasiclassical formalism and the experimentally observed internal phase transition line in $\mathrm{CeRh_2As_2}$ is likely caused by physical effects not included in the present model, such as the anisotropy of the material, coexisting orders, and electron correlations (see also the discussion in Sec.~\ref{sec:conclusion}).
Nevertheless, we emphasize that the limitation of the "approximation solution" also plays a non-negligible role at low temperatures.
Regarding this point, it is difficult to evaluate the temperature region where the "approximate solution" remains quantitatively reliable unless we obtain the exact solution.
The quantitative validity of the approximation was examined in a previous study~\cite{adachi2006basal} for single-band $s$-wave and $d$-wave superconductors.
In their analysis, it was found to reproduce the exact results with reasonably good accuracy for the superconducting free energy and the amplitude of the lowest Landau level component, at least down to $T/T_{\mathrm{c}}=0.5$. 
It is not clear whether the same conclusion applies in the present case, since the system considered in their study differs from ours; in particular, our model involves multiple bands and explicitly includes the paramagnetic depairing effect.
However, in our self-consistent iterative procedure for solving the Eilenberger equation, the convergence rate remains almost unchanged from the high-temperature region (e.g., $T/T_{\mathrm{c}0}^{e}=0.7$) down to about $T/T_{\mathrm{c}0}^{e}=0.3$ and $0.35$ for the BCS and PDW states, respectively. 
Although a convergence rate does not, by itself, ensure quantitative accuracy of the results, we expect that the approximation remains valid down to these temperatures.

\end{document}